\documentclass[twocolumn]{aastex631}

\usepackage{amsmath}
\usepackage{bm}
\usepackage{rotating} 

\usepackage{tikz-cd}
\tikzset{smalltext/.style={"\textup{\scriptsize #1}" description}}

\usepackage{esint}

\newcommand{\Rg}{R_{\mathrm{g}}}

\newcommand{\md}{\mathrm{d}}

\newcommand{\me}{\mathrm{e}}
\newcommand{\mi}{\mathrm{i}}

\newcommand{\bJ}{\bm{J}}

\newcommand{\br}{\bm{r}}
\newcommand{\bk}{\bm{k}}

\newcommand{\btheta}{\bm{\theta}}
\newcommand{\bn}{\bm{n}}

\newcommand{\bOm}{\bm{\Omega}}
\newcommand{\nn}{\nonumber}

\newcommand{\p}{\partial}

\newcommand{\mcE}{\mathcal{E}}

\usepackage[normalem]{ulem}

\newcommand\CHrem{\bgroup\markoverwith{\textcolor{magenta}{\rule[0.5ex]{2pt}{0.9pt}}}\ULon}

\begin{document}

\title{Galactokinetics II: Spiral structure}

\author[0000-0002-5861-5687]{Chris Hamilton}
\affiliation{Institute for Advanced Study, Einstein Drive, Princeton NJ 08540}

\author[0000-0002-8532-827X]{Shaunak Modak}
\affiliation{Department of Astrophysical Sciences, Princeton University, 4 Ivy Lane, Princeton NJ 08544}

\author[0000-0002-0278-7180]{Scott Tremaine}
\affiliation{Institute for Advanced Study, Einstein Drive, Princeton NJ 08540}

\correspondingauthor{Chris Hamilton} \email{chamilton@ias.edu}

\begin{abstract}
We present a unified theory of {linear} spiral structure in stellar disks. We begin by identifying the characteristic scales involved in the spiral structure problem and listing some quantitative requirements of a successful theory. We then write down the general linear response theory for thin disks, making clear the equivalence between different {representations} (e.g., Volterra, Landau, van Kampen) of the theory. Next, using the asymptotic expansions developed in our previous \textit{galactokinetics} paper, we consider spiral structure on different spatial scales and thereby show how several classic results --- including Lindblad-Kalnajs density waves, swing amplification, Lin-Shu-Kalnajs modes, and groove instabilities --- emerge as limiting cases. In addition, many of our asymptotic results connect smoothly when extrapolated to intermediate regimes, rendering the analytic theory valid over a larger range of scales than naively expected. Finally, we identify situations in which {nonlinear} physics is unavoidable. Though many nonlinear questions remain unanswered, we hope that the theoretical synthesis developed here will allow us to both connect and {distinguish} the plethora of ideas that have accumulated over the last six decades of spiral structure studies, and will provide a foundation upon which a comprehensive theory might ultimately be built.
\end{abstract}



\section{Introduction}
\label{sec:Introduction}

Spiral galaxies are some of the most beautiful objects in the Universe. They comprise the majority of galaxies at low redshift \citep{SellwoodMasters}, and may even form a significant minority out to redshifts as high as $z\simeq 3$ \citep{kuhn2024jwst}. Yet after many decades of astronomers' preoccupation with these objects, we still do not have a complete understanding of the structures that give spiral galaxies their name.

This shortcoming is significant for several reasons, but perhaps most important is that spirals are likely to be a major source of angular-momentum transport in galaxies. As a result, they drive secular evolution \citep{Kormendy2004} by churning the distribution of stellar orbits in galactic disks \citep{sellwood2002radial}, and also transport gas toward galaxies' central regions \citep{hopkins2011analytic}, feeding accretion onto supermassive black holes. Thus, spirals are responsible for a large amount of the redistribution of matter and feedback cycles that determine so much of galaxy formation and evolution. Without a complete, quantitative theory of the generation, maintenance, and characteristics of spiral structure, we cannot be confident in our understanding of how galactic machines work.

The \textit{theory} of spiral structure in galaxies presents a paradox. On the one hand, it is a jewel in the crown of theoretical galactic dynamics --- perhaps in no other sub-genre of the field has the level of analytical and numerical sophistication been so high. Accordingly, over the last half-century, spiral structure has been the focus of at least twelve dedicated review articles/book chapters/monographs \citep{marochnik1974structure,dekker1976spiral,rohlfs1977lectures,toomre1977theories,athanassoula1984spiral,bertin1996spiral,Binney2008-ou,dobbs2014dawes,bertin2014dynamics,shu2016six, seigar2017spiral,SellwoodMasters}. On the other hand, the subject can seem rather unwieldy: a disjointed collection of observational facts and ingenious theoretical ideas, cloaked in an obscure jargon. And, while the aforementioned references do an excellent job of explaining individual spiral-generating mechanisms, they tend to take a historical approach\footnote{In fact even the \textit{history} of this subject is contentious \citep{pasha2004density,pasha2004density2,marochnik2005west}.} to reviewing the literature rather than offering a theoretical synthesis. For instance, one is still left wondering: what is the precise connection between the two-armed kinematic density waves of Lindblad and \cite{kalnajs1973spiral}, the Lin-Shu-Kalnajs (LSK) dispersion relation for tightly wound waves \citep{lin1966spiral,Kalnajs1965}, the local shearing sheet/swing amplification results of \cite{goldreich1965ii} and \cite{julian1966non}, the various WKB-based analyses and feedback cycles of, e.g., \cite{Mark1974-jj,Mark1976-pz,bertin2014dynamics}, and 
the fully global approaches taken by, e.g., \cite{Kalnajs1976-gg,jalali2005unstable,De_Rijcke2019-uo,petersen2023predicting}? 
Are these mechanisms truly distinct, or are they merely different aspects of the same thing? And if they are distinct, under which circumstances --- and  on which length and time scales --- should we expect a given mechanism to dominate?

The purpose of this paper is to build up a unified picture of \textit{linear} spiral structure in stellar disks from first principles, starting from just a few basic observational requirements. Our main assumptions are that the galactic disk in question is not too far from axisymmetric, that nonlinear coupling between fluctuations is unimportant, and that the spiral structure is dominated by the stars. These assumptions are sufficient to develop a theory in which many answers to the above questions emerge naturally, with almost all the classic (linear) results arising as particular asymptotic limits. More precisely, these are often the limiting regimes of spiral \textit{wavelength}, meaning that we can make use of the \textit{galactokinetic} formalism that we developed in \citet{Hamilton2024} (hereafter Paper I). There we showed that one could classify gravitational potential fluctuations by wavelength relative to a stellar orbit's guiding center radius $R_\mathrm{g}$ and epicyclic amplitude $a_R$. Not only did the analysis of those fluctuations simplify greatly in the asymptotically long- and short-wavelength regimes, but the asymptotic results also connected smoothly when extrapolated to intermediate wavelengths. Here, we apply the insights gained in Paper I to the problem of self-consistent spiral structure, constructing a theory that unifies many classic ideas, but also generalizes them.


Of course, reality is nonlinear. On its own, linear theory is incapable of answering questions about how spiral instabilities saturate \citep{sellwood2022spiral} or how they give rise to further instabilities \citep{sellwood2019spiral}, how spiral responses can self-perpetuate \citep{d2013self,Sellwood2020-ao}, and how two or more distinct spiral modes can couple \citep{Tagger1987-sf}. These fundamentally nonlinear questions are the focus of current theoretical research \citep{fouvry2015secular,Sridhar2019-zo,hamilton2024saturation,roule2025long}, but are beyond the scope of this paper (though see \S\ref{sec:Nonlinear}). However, many important questions \textit{can} be answered with linear theory alone, and in fact most nonlinear studies do rely on linear theory to some extent for the construction and/or interpretation of their results. 
Similarly, we know that gas is a crucial ingredient for both the formation and maintainence spiral structures (e.g., \citealt{Sellwood1984-bw}), and that there can be non-trivial coupling between gaseous and stellar spirals \citep{rafikov2001local,kim2007gravitational,george2025redefining}, none of which we address carefully here.
Thus, a robust linear theory of spirals in stellar disks --- the topic of this paper --- is a necessary, but certainly not sufficient, condition for a proper understanding of spiral structure as a whole. 

The remainder of this paper is organized as follows. In \S\ref{sec:Overview} we provide a very brief overview of the spiral structure problem, identifying some necessary criteria for a successful theory, and introducing the characteristic scales and simplifying assumptions that will allow us to build up a coherent linear theory from first principles. In \S\ref{sec:Linear} we introduce the formal apparatus of linear response theory for thin stellar disks. We then proceed to investigate the linear theory in several limiting regimes. First is the limit of long wavelengths (\S\ref{sec:Long_Wavelengths}), where we recover Lindblad-Kalnajs kinematic density waves and gain insight into global instabilities, including a new derivation of the groove instability. Second is the limit of short wavelengths (\S\ref{sec:Short_Wavelengths}), leading to a rederivation --- and generalization --- of swing amplification, and recovery of the LSK dispersion relation. In the Discussion (\S\ref{sec:Discussion}) we explain how the spiral structure literature of the past fits into our scheme, and discuss some open questions in (primarily nonlinear) spiral structure theory. We summarize in \S\ref{sec:Summary}.

Note that throughout this paper, we employ the same notation as in Paper I, sometimes without repeating the definition.


\section{Overview of the problem}
\label{sec:Overview}




In this section, we first define spiral geometry and briefly discuss the observed morphology (\S\ref{sec:geometry}) of spiral arms. We then comment on the {amplitudes} of spirals and what this implies about amplification mechanisms (\S\ref{sec:amplitude_requirements}), before gathering characteristic length and time scales involved in the spiral structure problem (\S\ref{sec:Characteristic_Scales}). Finally, we justify the key simplifying assumptions we will make throughout the rest of the paper (\S\ref{sec:Simplifying_Assumptions}).

\subsection{Spiral geometry and morphology}
\label{sec:geometry}

An arbitrary non-axisymmetric potential fluctuation in a 2D disk described by polar coordinates $(\varphi, R)$ can be written
\begin{equation}
    \delta \phi(\br) = \sum_{m=1}^\infty S_m(R)\cos[m\varphi+f_m(R)],
\end{equation}
for some real functions $S_m(R)$ and $f_m(R)$. Then `spirals' are just those fluctuations for which a single $m$ dominates this sum, and for which $f_m$ is monotonic in $R$. Dropping the $m$ subscripts, the trough of a spiral pattern's potential can be written
\begin{equation}
    m\varphi + f(R) = \,\mathrm{const.} \,\,\,\,\,\,\,\,\, \mathrm{(mod \,\,2\pi)},
    \label{eqn:R_of_phi}
\end{equation}
where $m>0$, and $f(R) \equiv mg(R)$ is called the \textit{shape function}. The pitch angle $\alpha \in (0,\pi)$ of this spiral is defined through
\begin{equation}
    \cot \alpha =  -R\frac{\p \varphi}{\p R} = R \frac{\md g}{\md R}.
    \label{eqn:def_pitch}
\end{equation}
We always take the net rotation of the disk to be in the positive $\varphi$ direction; then trailing arms have $\md g/\md R > 0$ (i.e., pitch angles $\alpha\in (0,\pi/2)$) while leading arms have $\md g/\md R < 0$ (i.e., pitch angles $\alpha\in (\pi/2,\pi)$). 

The $n_\varphi$th azimuthal Fourier component of the $m$-armed spiral fluctuation --- see equation (61) of Paper I --- is
\begin{equation}
    \delta \phi_{n_\varphi}(R) = \frac{1}{2}S(R)\me^{in_\varphi g(R)},
\end{equation}
for $n_\varphi = \pm m$, and zero otherwise. Following the definitions from \S4.1 of Paper I, the corresponding wavevector at guiding radius $R_\mathrm{g}$ is $\bk = (k_\varphi, k_R)$, where
\begin{align}
    k_\varphi &= \frac{ n_\varphi }{R_\mathrm{g}}, \\
    k_R &= \left(-i \frac{\md \ln S}{\md R} + n_\varphi \frac{\md g}{\md R}\right)_{R=R_\mathrm{g}} \nn \\
    &= \frac{n_\varphi \cot \alpha}{R_\mathrm{g}}\left(1-i \frac{\md \ln S}{\md \ln R} \frac{\tan \alpha}{n_\varphi} \right)_{R=R_\mathrm{g}},
    \label{eqn:spiral_radial_wavenumber}
\end{align}
and to get the last line we used \eqref{eqn:def_pitch}. In the common scenario where the imaginary term in the bracket in \eqref{eqn:spiral_radial_wavenumber} is much smaller than unity, we can approximate 
\begin{equation}
    k \simeq \bigg \vert \frac{n_\varphi}{R_\mathrm{g}\sin\alpha}\bigg\vert,
\end{equation}
and hence estimate
\begin{align}
    k a  \simeq {0.9}\,\times   &  \left(\frac{m}{2}\right)   \left(\frac{\sin \alpha}{0.2}\right)^{-1}  \left(\frac{\gamma}{\sqrt{2}}\right)   \nn \\
    &\times \left(\frac{\sigma}{20\,\mathrm{km\,s}^{-1}}\right)\left(\frac{V}{220\,\mathrm{km\,s}^{-1}}\right)^{-1},
    \label{eqn:ka_estimate}
\end{align}
where $a$ is the rms epicyclic amplitude, $\sigma$ is the radial velocity dispersion, $V$ is the circular orbital speed, and the frequency ratio $\gamma \equiv 2\Omega/\kappa$ is equal to $\sqrt{2}$ for a flat rotation curve.

Real galactic spirals can have quite a wide array of morphologies, i.e., a variety of amplitudes, arm numbers $m$, and pitch angles $\alpha$. They span a range from two-armed `grand design' spirals like M51, through few-armed `intermediate' scale spirals like M61 (and, presumably, the Milky Way), to many-armed `flocculent' spirals like NGC\,4414 ---  see, e.g., \citet{Binney2008-ou,yu2020statistical,SellwoodMasters} and references therein. Also, spiral morphologies can differ depending on the wavelength of observation, reflecting the fact that, e.g., gaseous and stellar spirals can arise via distinct mechanisms.
In fact the gaseous interstellar medium (ISM) can play an important role in the dynamics of purely \textit{stellar} spirals, both as a source of new stars \citep{Sellwood1984-bw} and as a bubbling sea of gravitational potential fluctuations \citep{modak2025characterizing}.
As mentioned in the Introduction, in this paper we consider spirals in the stellar component of galactic disks; the ISM is present only implicitly, as one possible source of noise.

Despite the variety of observed spiral structure, some quantitative trends are apparent \citep{yu2020statistical}.  There is a strong preference for $m=2$ armed, trailing spirals, nearly always with pitch angles in the range $\alpha\in(5^\circ, 40^\circ)$, i.e., $1 \lesssim \cot \alpha < 10$. Plugging this range into the estimate \eqref{eqn:ka_estimate} for the fiducial values given there, we get $ka \in (0.3, 2.3)$. In fact, more than $90\%$ of nearby spirals have $\alpha\in(10^\circ, 30^\circ)$, i.e., $1.7 \lesssim \cot \alpha < 5.7$, which tightens the estimate \eqref{eqn:ka_estimate} to $ka \in (0.4, 1.2)$.

One could further imagine classifying spiral morphology according to the mass, star formation rate, galactic environment, etc. of their host galaxies \citep{yu2020statistical}. The ultimate theory of spiral structure would successfully predict the statistics of the spiral morphology as a function of these host properties. Unfortunately, we are not close to possessing such a theory. A more modest goal is to explain, at least post-hoc, why a galaxy of a given type and in a given (either real or simulated) environment has the spiral structure that it does, how long that spiral is likely to last, and what the mechanism behind its generation and maintenance is. The linear theory of spiral structure can take us a long way toward achieving this goal.

\subsection{Spiral amplitudes}
\label{sec:amplitude_requirements}

The majority of observed star-dominated spiral arms have amplitudes (over/underdensities $\delta \Sigma$) of a few percent to a few tens of percent relative to the background disk surface density $\Sigma$ \citep{yu2020statistical,smith2022effect}:
\begin{equation}
    \frac{\vert \delta \Sigma \vert}{\Sigma} \sim 0.1 \,\,\,\,\,\,\,\,\,\,\,\,\,\,\mathrm{(observed).}
    \label{eqn:amplitude_observed}
\end{equation}
This is much larger than the typical level of background noise in the disk would imply. It follows that many spirals must be strongly \textit{amplified}.

To get an idea of how much amplification is required, consider the main sources of noise, $\delta \phi^\mathrm{ext}$, in the disk. First there are the inevitable potential fluctuations driven by the turbulent ISM, at the  level of say $\delta \phi^\mathrm{ext} \lesssim (6\,\mathrm{km\,s}^{-1})^2$ \citep{modak2025characterizing}. In the absence of any amplification, linear theory would suggest that these drive fluctuations $\vert\delta \Sigma\vert /\Sigma \sim \vert \delta\phi^\mathrm{ext}\vert/V^2$. Supposing the potential fluctuations are instead amplified by a dimensionless factor $\sim \mathcal{A}$, the resulting surface-density fluctuations would be at the level 
\begin{align}
    &\frac{\vert \delta \Sigma \vert}{\Sigma} \sim 0.1 \times  \left( \frac{\mathcal{A}}{10^2}\right) \left( \frac{\vert \delta \phi^\mathrm{ext} \vert }{(6\,\mathrm{km\,s}^{-1})^2} \right) \left( \frac{V}{220 \,\mathrm{km\,s}^{-1}} \right)^{-2}.
    \label{eqn:amplitude_vs_observation}
\end{align}
Comparing with \eqref{eqn:amplitude_observed}, we see that amplification factors of $\mathcal{A} \sim 10^2$ would be necessary for a spiral seeded by these ISM fluctuations to reach the required amplitude.

A very similar argument applies to spirals seeded by a spectrum of substructure in a galaxy's dark halo. \cite{gilman2025dark} noted that realistic subhalo models produce fluctuations $\vert \delta \phi^\mathrm{ext} \vert \lesssim$ ($1\mathrm{\,km\,s}^{-1}$)$^2$ in a Milky-Way-like galaxy. From \eqref{eqn:amplitude_vs_observation}, these would need to be amplified by $\mathcal{A}\sim 10^3$ to match the observed spirals.
A different type of `noise' that could seed spiral structure is the infall of a dwarf galaxy like Sagittarius \citep{bland2021galactic}. Such an encounter can realistically provide $\vert \delta \phi^\mathrm{ext} \vert \sim (20\,\mathrm{km\,s}^{-1})^2$, which would still likely require some amplification, say $\mathcal{A}\sim 10$. There are exceptional cases like M51, in which there has been a recent encounter with a galaxy of similar size, meaning little or no self-gravitating amplification of spiral structure is required. However, it is clear that not all spirals can be produced exclusively in this way because such encounters are rare, and many galaxies with significant spiral amplitudes do not have obvious companions. Similar considerations apply to spiral structure driven by a strong bar, like in NGC\,1300.


Thus, we expect that in order to understand the amplitudes of spirals, especially (but not only) in galaxies \textit{without} significant perturbers like infalling dwarfs or a strong bar, we need to account for the amplification of the typical noise level by factors of at least $10$ or $100$. This requirement places a strong filter on the portions of parameter space relevant for our theory: we can often ignore regions of parameter space where self-gravity is formally strong, if it only leads to, say, $\mathcal{A}\sim 3$.

\subsection{Characteristic time and length scales}
\label{sec:Characteristic_Scales}

The generation of spiral structure is, in a superficial way, a rather trivial problem: any  passive tracer added to a differentially-rotating flow will inevitably shear into a spiral pattern, due to the radially-dependent circular frequency profile $\Omega(R)$. Thus, at face value the fact that many galactic disks look spirally is not surprising.

However, this idea immediately gives rise to a \textit{winding problem}. Suppose our tracer consisted of stars or gas on circular orbits --- a so-called `material arm' --- arranged into a blob of size $\sim l$. Then the bits of the blob at different radii will contain stars whose azimuthal velocities differ  by $\delta v_\varphi \sim R\times (l \times \md \Omega / \md R) \sim l \Omega$, and as such will separate by a distance $\sim l$ on a `shear' timescale
\begin{align}
    t_\mathrm{shear} \sim  \frac{l}{\delta v_\varphi} \sim \frac{1}{\Omega},
    \label{eqn:tshear}
\end{align}
which is independent of $l$, and comparable to the circular orbital period (much shorter than the age of a galaxy).
If all spirals were material arms, many of them would be far more tightly wound ($\alpha \to 0$) than is observed (see, e.g., \S6.1.3 of \citealt{Binney2008-ou}). 
To construct a theory capable of solving the aforementioned problems, we will have to drop the assumption of circular orbits, or drop the assumption of passive tracers, or both. Each of these will introduce new characteristic scales into the spiral structure problem.

First let us drop the assumption of passive tracers, and allow that inhomogeneities in the disk generate their own gravity.
Then we might guess that the (amplified) spirals are the manifestation of some self-gravitating effect, like an instability. An overdense blob of matter of size $\sim l$ in this disk wants to contract under its own gravity on a `collapse' timescale
\begin{align}
    t_\mathrm{coll}(l) \sim \sqrt{\frac{l}{G\Sigma}}.
    \label{eqn:tcoll}
\end{align}
Ignoring non-circular motions for now, this contraction will still have to compete with the background shear: the shear will prevent contraction if $t_\mathrm{shear}$ (equation \eqref{eqn:tshear}) is $ \lesssim t_\mathrm{coll}$, or rather
$l \gtrsim l_\mathrm{shear}$ where
\begin{equation}
    l_\mathrm{shear} \sim \frac{G\Sigma}{\Omega^2}.
    \label{eqn:lshear}
\end{equation}
Now, let us account for the fact that (stellar) orbits are not circular but are rather epicyclic.
At a given $R_\mathrm{g}$, the majority of stars will have radial velocities $\sim \sigma$, so the blob will be rearranged on a `dispersion' timescale
\begin{align}
    t_\mathrm{dis}(l) \sim \frac{l}{\sigma}.
    \label{eqn:tdis}
\end{align}
The self-gravitating collapse of our blob will then be halted by 
dispersion if $t_\mathrm{dis} \lesssim t_\mathrm{coll}$, or, using equations \eqref{eqn:tcoll} and \eqref{eqn:tdis}, equivalently $l \lesssim  l_\mathrm{dis}$ where
\begin{equation}
     l_\mathrm{dis} \sim  \frac{\sigma^2}{G\Sigma}.
    \label{eqn:ldis}
\end{equation}

Putting these estimates together, we expect that if  $ l_\mathrm{shear} \lesssim  l_\mathrm{dis}$, blobs on all scales will be stable, whereas if $ l_\mathrm{shear} >  l_\mathrm{dis}$ there should exist a range of length scales $ l_\mathrm{dis} \lesssim  l \lesssim  l_\mathrm{shear}$ for which instability is possible. Comparing equations \eqref{eqn:lshear} and \eqref{eqn:ldis} we find that for instability to be possible we require $\sigma \Omega / (G\Sigma) \lesssim 1$.  Anticipating a famous result of linear response theory \citep{toomre1964gravitational}, we hereafter refine this criterion to
\begin{equation}
    Q \lesssim 1,
\end{equation}
where
\begin{equation}
    Q \equiv \frac{\sigma \kappa}{3.36 \,G\Sigma },
    \label{eqn:Q_Toomre}
\end{equation} 
is Toomre's $Q$ parameter.

These estimates suggest that we use the Toomre $Q$ parameter as a proxy for when we should expect interesting self-gravitating spiral dynamics to occur, and especially when we should expect spiral perturbations to be amplified significantly. If $Q \gg 1$ the disk is very stable on all scales, and so self-gravitating effects will be suppressed. On the other hand, if $Q \ll 1$, the disk would be horribly unstable and immediately collapse on a dynamical timescale, probably settling into something with a larger $Q$. When $Q \sim 1$, we expect that our disk is either weakly stable or weakly unstable (or indeed marginally stable).
Returning to equations \eqref{eqn:lshear} and \eqref{eqn:ldis}, whenever $Q\sim 1$ we have 
\begin{equation}
     l_\mathrm{shear}\sim l_\mathrm{dis} \sim \frac{\sigma}{\kappa} \sim a,
\end{equation}
the typical epicyclic amplitude, and for fluctuations on this scale we have
\begin{equation}
    t_\mathrm{coll}(a) \sim t_\mathrm{dis}(a) \sim \frac{1}{\Omega} \sim t_\mathrm{shear},
    \label{eqn:all_the_times}
\end{equation}
comparable to the typical orbital time.
Thus, we expect that most relevant self-gravitating spiral phenomena will occur when Toomre's $Q\sim 1$, on length scales $ka\sim 1$ (consistent with the observational estimates made in \S\ref{sec:geometry}), and will be generated and/or destroyed on timescales not much shorter than $\sim \Omega^{-1}$.

\subsection{Simplifying assumptions}
\label{sec:Simplifying_Assumptions}

The key assumptions we will make throughout this paper are that (i) we may consider spirals as linear perturbations of an otherwise axisymmetric galactic disk, (ii) vertical dynamics are unimportant so the problem may be considered effectively two-dimensional, and (iii) the spiral/disk dynamics do not backreact upon the galaxy's host dark matter halo (although the substructure in the halo may well perturb the disk). Since so much follows from a theory based upon these assumptions, it is worth justifying them individually.

(i) \textit{Linear perturbations}. Since most spiral over/underdensities are $\lesssim 10\%$ (\S\ref{sec:amplitude_requirements}), the corresponding potential fluctuations will be similarly small (especially since a significant component of the background potential is set by hot components of the galaxy like the dark halo). Hence, the potentials of spiral-hosting stellar disks are not too far from axisymmetric\footnote{The major exceptions to this rule are to be found in barred galaxies, in the region close to or inside the bar, although if we are interested in spiral structure at radii well beyond the bar radius we can still treat the background disk as axisymmetric and the bar as a persistent, weak, non-axisymmetric driving perturbation.}, meaning the orbits of stars should be very nearly integrable and describable using angle-action variables \citep{Binney2013-cf}. Then on short enough timescales, the orbital motion can be split into a dominant `mean field' piece driven by the axisymmetric component of the gravitational potential, plus a weak (linear) perturbation due to the spirals and any other non-axisymmetric fluctuations. We discuss the nonlinear breakdown of this assumption in \S\ref{sec:Nonlinear}; for now it suffices to say that
even the emergence of nonlinear structures will almost certainly involve, at least early on, a significant period of linear evolution. Linear theory is, therefore, an absolutely necessary (though insufficient) component of spiral structure theory.





(ii) \textit{Ignoring vertical motion}. In this case observations are less helpful, but one can sensibly argue that 
the timescales associated with spirals will be those tied to azimuthal and radial motions in the disk, and since the vertical frequency tends to be much higher than both the radial and azimuthal frequencies, vertical motions should be largely adiabatically invariant under spiral perturbations \citep{Binney1988-zy}. Moreover, stellar populations with significant vertical actions tend also to have significant radial actions (at least in the Milky Way; see \citealt{Mackereth2019-cq}), so we would not expect them to play a major role even in our effective 2D theory.

(iii) \textit{Frozen halo}. We know that galaxies are embedded in dark haloes, and indeed substructure in the halo might excite spiral waves in the stellar disk (\S\ref{sec:amplitude_requirements}). On the other hand, we need to know whether to account for any backreaction of the spirals onto the halo --- in other words whether the halo needs to be `live'  in our calculations, or can instead be idealized as frozen. Theoretical calculations by \cite{Mark1976-xx} and \cite{fuchs2004density} have suggested that dynamics in the presence of a live halo might be quite different to those with a frozen halo. However, the only author to have tested this claim in published global simulations seems to be \cite{Sellwood2021-rt}, who found that there was little spiral-halo coupling at all, at least in the context of groove instabilities (\S\ref{sec:Long_Wavelength_Instability_Sharp}). It seems that this question is not fully settled --- in particular, what physics accounts for the fact that galactic \textit{bars} certainly couple strongly to the halo \citep{Tremaine1984-wt,athanassoula1984spiral,Chiba2021-cc} if (at least some) spirals do not? In lieu of further investigations, we follow the conclusion of \cite{Sellwood2021-rt} that a frozen halo is good enough for our purposes.


\section{Linear theory}
\label{sec:Linear}

The arguments of \S\ref{sec:Overview} suggest that we construct a theory of spiral structure based upon the linear response properties of a razor-thin stellar disk, which is embedded in a fixed halo potential and possibly subject to external potential fluctuations. We write down the formal equations of this linear  theory in \S\ref{sec:General_Linear}. In \S\ref{sec:choice_of_representation} we briefly mention the issue of representation degeneracy in linear theory. We opt for the Volterra representation, whose central object is the Volterra kernel \eqref{eqn:Volterra_Kernel}. In \S\ref{sec:Simplifying_Volterra} we discuss the interpretation of this kernel and provide a some numerical examples.

\subsection{General equations}
\label{sec:General_Linear}

The distribution function (DF) of our disk can be expressed in angle-action variables as $f(\btheta,\bJ,t) = f_0(\bJ, t) + \delta f(\btheta, \bJ, t)$, where $f_0$ is the angle-averaged DF. In linear theory we treat $f_0 = f_0(\bJ)$ as independent of time and consider the evolution of the (presumed small) fluctuation $\delta f$. This approximation is valid for times $t\ll t_\mathrm{NL}$, where $t_\mathrm{NL}$ is some nonlinear timescale that we discuss in \S\ref{sec:Nonlinear}.

Taking equation (19) of Paper I and discarding nonlinear terms we have
\begin{align}
    \label{eq:Fluctuation_Evolution}
    &\frac{\partial \delta f}{\partial t} + \bOm \cdot \frac{\partial \delta f}{\partial \btheta} - \frac{\partial f_0}{\partial \bJ} \cdot \frac{\partial \delta \phi^\mathrm{tot}}{\partial \btheta} = 0,
\end{align}
where  $\delta \phi^\mathrm{tot}$ is the full potential fluctuation,
\begin{equation}
    \delta \phi^\mathrm{tot} = \delta \phi + \delta \phi^\mathrm{ext}.
    \label{eqn:total_potential_fluctuation}
\end{equation}
Here $\delta \phi$ is the part of the potential fluctuation sourced directly by $\delta f$:
\begin{equation}
    \nabla^2 \delta \phi = 4\pi G \int \md \bm{v} \, \delta f,
    \label{eqn:poisson_fluctuating}
\end{equation}
and $\delta \phi^\mathrm{ext}$ is any additional, externally imposed piece. Expressing \eqref{eq:Fluctuation_Evolution} as a Fourier series in angles and integrating in time gives
\begin{align}
    \label{eq:linear_Vlasov_formal_solution}
    &\delta f_{{\bm{n}}}(\bm{J}, t) =  \delta f_{{\bm{n}}}(\bm{J}, 0) \me^{-i\bn\cdot\bOm t} 
    \nn
    \\
    &\,\,\,\,\,\,\,\,\,\,\,\,\,+  i {\bm{n}}\cdot \frac{\p f_0}{\p \bm{J}}\int_0^t \md t' \, \me^{- i {\bm{n}}\cdot \bOm (t-t')} \delta\phi^\mathrm{tot}_{{\bm{n}}}(\bm{J}, t').
\end{align}

Equations \eqref{eqn:total_potential_fluctuation}-\eqref{eq:linear_Vlasov_formal_solution} constitute our linear response problem; they now need to be solved simultaneously. The standard method for achieving this is the \textit{biorthogonal basis method} \citep{Kalnajs1976-gg,Binney2008-ou,hamilton2024kinetic}, which involves projecting potential and surface density fluctuations onto a complete set of (generally complex) basis functions ${ (\phi^{(p)} (\br) , \Sigma^{(p)} (\br)) }$ that satisfy
\begin{align}
    & \phi^{(p)} (\br) = \!\! \int \!\! \md \br' \, \psi (\br , \br') \, \Sigma^{(p)} (\br') ,
    \label{def_basis_1}
    \\
    & \!\! \int \!\! \md \br \, [\phi^{(p)}(\br)]^* \, \Sigma^{(p')} (\br) = - \mcE \delta_{p}^{p'} ,
    \label{def_basis_2}
\end{align}
where
\begin{equation}
    \psi (\br , \br') = - \frac{G}{\vert \br - \br' \vert },
    \label{eqn:Newtonian_Kernel}
\end{equation}
and $\mcE$ is an arbitrary (real) normalization factor with units of energy.
One can then expand the potential fluctuations as
\begin{eqnarray}
    \delta \phi(\br, t) &=& \sum_p B^p(t) \phi^{(p)}(\br),
    \label{eqn:phi_basis}
    \\
    \delta \phi^\mathrm{ext}(\br, t) &=& \sum_p B_\mathrm{ext}^p(t) \phi^{(p)}(\br).
    \label{eqn:phiext_basis}
\end{eqnarray}
for some (unknown) coefficients $B^p(t)$ and (known) coefficients $B_\mathrm{ext}^p$.
In Fourier language these expansions read
\begin{eqnarray}
    \delta \phi_{\bn}(\bJ, t) = \sum_p B^p(t) \phi_{\bn}^{(p)}(\bJ),
    \label{eqn:phi_basis_Fourier}
    \\
    \delta \phi_{\bn}^\mathrm{ext}(\bJ, t) = \sum_p B_\mathrm{ext}^p(t) \phi_{\bn}^{(p)}(\bJ).
    \label{eqn:phiext_basis_Fourier}
\end{eqnarray}
The surface density fluctuations can be expanded similarly:
\begin{align}
 \delta \Sigma(\br, t) &= \sum_p B^p(t) \Sigma^{(p)}(\br),
    \label{eqn:DeltaSigma_basis}
    \\
    \delta \Sigma_{\bn}(\bJ, t) &= \sum_p B^p(t) \Sigma_{\bn}^{(p)}(\bJ).
\end{align}
Most importantly, if we can solve for the set of coefficients $\{B^p(t)\}$, we can then reconstruct the potential fluctuations using \eqref{eqn:phi_basis_Fourier}-\eqref{eqn:phiext_basis_Fourier}, and reconstruct the fluctuation in the DF by substituting the result back into \eqref{eq:linear_Vlasov_formal_solution}.

Before we write down a closed equation for $B^p$, we mention two useful results that follow from the definitions above. First, from \eqref{def_basis_1}-\eqref{def_basis_2} that we can expand the Newtonian kernel \eqref{eqn:Newtonian_Kernel} in terms of basis functions:
\begin{align}
    \psi(\br,\br')&=-\frac{1}{\mathcal{E}} \sum_p \phi^{(p)}(\br)[\phi^{(p)}(\br')]^* \\
    &=\sum_{\bn, \bn'}\psi_{\bn\bn'}(\bJ,\bJ')\me^{i(\bn\cdot\btheta-\bn'\cdot\btheta')},
    \label{eqn:kernel_expansion}
\end{align}
where in the second line we converted to angle-action coordinates, and defined
\begin{equation}
    \psi_{{\bn}{\bn}'} (\bJ , \bJ')  \equiv - \frac{1}{\mcE}\sum_{p} \phi^{(p)}_{{\bn}} (\bJ) \, [\phi^{(p)}_{{\bn}'} (\bJ')]^*.
    \label{bare_basis_app}
\end{equation}
Second, we can get a formal expression for $B^p$ by multiplying \eqref{eqn:DeltaSigma_basis} by $[\phi^{(p)}(\br)]^*$, integrating over $\br$, and using the orthogonality property \eqref{def_basis_2} and the fact that $\delta \Sigma = \int \md \br \,\delta f$. The result is
\begin{align}
       B^p(t) &= -\frac{1}{\mathcal{E}} \int \md \br \, \md \bm{v} \, [\phi^{(p)}(\br)]^*\delta f (\br, \bm{v}, t),
       \\
       &= -\frac{(2\pi)^2}{\mathcal{E}}\sum_{\bn}\int \md \bJ \, [\phi^{(p)}_{\bn}(\bJ)]^*\delta f_{\bn}(\bJ, t),
       \label{eqn:B_formal}
\end{align}
where to get the second line we used the fact that angle-action coordinates are canonical, $\md \br \, \md \bm{v} = \md \btheta \, \md \bJ$.

Now we can plug the expansions \eqref{eqn:phi_basis_Fourier}-\eqref{eqn:phiext_basis_Fourier} into the right hand side of \eqref{eq:linear_Vlasov_formal_solution}, and insert the result into \eqref{eqn:B_formal}, thereby deriving a Volterra equation for the coefficients $B^p(t)$:
\begin{align}
    & B^p(t) = 
    B^p_\mathrm{kin}(t) \nn \\
    &+  \int_0^t \md t' \sum_{p'} \mathcal{M}^{pp'}(t-t') [B_\mathrm{ext}^{p'}(t') + B^{p'}(t')],
    \label{eqn:Volterra_Equation}
\end{align}
where
\begin{align}
    B^p_\mathrm{kin}(t) &= -\frac{(2\pi)^2}{\mathcal{E}}\sum_{\bn}\int \md \bJ [\phi^{(p)}_{\bn}(\bJ)]^*
    \delta f_{\bn}(\bJ, 0)\me^{-i\bn\cdot\bOm t}
    \label{eqn:B_kinematic}
\end{align}
arises from pure kinematic evolution of the initial fluctuation, and the Volterra kernel
\begin{align}
    \mathcal{M}^{pp'}(\tau) = -\frac{(2\pi)^2}{\mathcal{E}} & \sum_{{\bm{n}}}\int  \md \bm{J} \, i\bn\cdot \frac{\p f_0}{\p \bJ} \, \me^{ - \mi {\bm{n}} \cdot \bOm \tau } \nn \\
    & \times 
    [\phi^{(p)}_{\bm{n}} (\bm{J})]^*
    \phi^{(p')}_{\bm{n}} (\bm{J}),
    \label{eqn:Volterra_Kernel}
\end{align} 
which has units of frequency, encodes the response to potential fluctuations. The kernel \eqref{eqn:Volterra_Kernel} will be the main object of interest throughout much of this paper.

\subsection{A note on representations of linear theory}
\label{sec:choice_of_representation}

It is worth pausing to mention a technical issue, which is a source of confusion when reading and comparing the classic studies on spiral structure\footnote{For concreteness we continue to focus here exclusively on stellar disks, but similar issues occur for fluid disks \citep{lin1987spectral}.}.
The issue is that the solution to the above linear response theory can be formulated in (at least) four different ways, and the links between these representations are rarely made explicit.  Because of this `representation freedom,' it is often the case that two or more theoretical approaches can appear totally distinct when they in fact contain very similar physical content. To be clear, we are not saying that the choice of representation is the \textit{only} difference between theories of spiral structure --- there are genuinely important physical differences too. But it is important to be clear about which aspects of spiral structure theory only appear different because of alternative choices of representation. 

The most direct approach, and the one we will use throughout most of this paper, is to work with the Volterra equation \eqref{eqn:Volterra_Equation}. We call this the \textit{Volterra representation}. It is convenient for numerical calculations since we work in the time domain and can solve \eqref{eqn:Volterra_Equation} iteratively (e.g., \citealt{Murali1999-eh,binney2020shearing,rozier2022constraining}). On the other hand, this approach does not make manifest the underlying `modal structure' of the system (by which we mean the fact that the system really does prefer a small subset of frequencies over all the others, e.g., \citealt{pichon1997numerical,evans1998stability1,evans1998stability2}). 
In Appendix \ref{sec:App_Representations} we sketch three alternative representations of linear theory, what we call the \textit{Landau representation} (in which the key objects are the potential and density fluctuations in complex frequency space), the van Kampen representation (which is closely related to the Landau approach but focuses primarily on real frequencies), and the Langrangian representation (which involves the trajectories of individual stars rather than the Eulerian potential or DF fields we are discussing here). We also give some references to where these alternative representations have appeared in important spiral-related literature in the past. See \S\ref{sec:previous_literature} for further discussion.

\subsection{The Volterra kernel}
\label{sec:Simplifying_Volterra}

The Volterra kernel \eqref{eqn:Volterra_Kernel} encodes the self-gravitating response of the disk to internal and/or external perturbations. More precisely, the function $\mathcal{M}^{pp'}(\tau)$ tells us the extent to which a fluctuation with spatial structure $\phi^{(p)}$ at time $t$ remembers a previous fluctuation with spatial structure $\phi^{(p')}$ from an earlier time $t-\tau$. Inspection of \eqref{eqn:Volterra_Kernel} suggests that when viewed as a function of $\tau$, the kernel $\mathcal{M}^{pp'}(\tau)$ will oscillate on a roughly orbital timescale with an overall decaying envelope (due to the fact that the integrand oscillates rapidly with $\bJ$ at large $\tau$). This decay corresponds to the system `forgetting' its own history.

For some systems self-gravity will be negligible. To make this statement quantitative, recall from Paper I that when analyzing an individual orbit we always ignored relative corrections $\mathcal{O}(\epsilon_R^2)$ to the unperturbed motion, where the small parameter $\epsilon_R \sim a_R/R_\mathrm{g} \lesssim 0.1$ measured the orbit's eccentricity. When performing calculations that involved a whole population of orbits we ignored relative corrections $\mathcal{O}(\epsilon^2)$, where $\epsilon \equiv \sqrt{\langle \epsilon_R^2 \rangle}$. Thus, throughout this paper we declare self-gravity negligible if it provides at most an $\mathcal{O}(\epsilon^2)$ relative correction to the potential/DF fluctuations.

To turn this into a rough criterion on $\mathcal{M}$, note that $\mathcal{M}^{pp'}(\tau)$ typically accumulates its maximum value by the time the phase $\bn\cdot\bOm\tau$ reaches $\simeq \pi/2$, which we approximate as $\tau \simeq \tau_{\pi/2}$, where
\begin{align}
    \tau_{\pi/2} &\equiv \frac{\pi/2}{m\Omega } = \frac{T_\varphi}{4m},
    \label{eqn:tau_pi_by_2}
\end{align}
with $\Omega$ the typical circular frequency. Then bare potential fluctuations will be amplified by self-gravity by a factor\footnote{This result is more familiar in the Landau representation (\S\ref{sec:Landau}), which tells us that, roughly speaking, a stimulus $\widehat{\delta \phi}(\omega)$ has its bare potential `dressed' according to the matrix $[\mathcal{I}-\widehat{\mathcal{M}}(\omega)]^{-1}$.}  $\sim \vert 1- \mathcal{M}^{pp'}(\tau_{\pi/2})\,\tau_{\pi/2}\vert^{-1}$.
Since in linear theory, potential fluctuations are imposed at a level $\vert \delta\phi/h_0\vert  \sim \epsilon$ (see \S5.1 of Paper I for justification), a condition for  self-gravitating (finite-$\mathcal{M}$) corrections to be ignorable ($\mathcal{O}(\epsilon^2)$ or smaller) is that 
\begin{equation}
   \vert  \mathcal{M}^{pp'}(\tau_{\pi/2}) \vert  \, \tau_{\pi/2} \lesssim  \epsilon.
    \label{eqn:small_kernel}
\end{equation}
Of course, since the envelope of $\mathcal{M}^{pp'}(\tau) $ must decay at late times, we will always have $\vert \mathcal{M}^{pp'}(\tau) \vert  \, \tau_{\pi/2} \lesssim  \epsilon$ for all $\tau$ bigger than some {convergence time} $\tau_\mathrm{conv}$, regardless of how strong self-gravity was at smaller values of  $\tau$. 

We emphasize that the inequality \eqref{eqn:small_kernel} is a very crude estimate for when self-gravity is negligible, and should not replace a proper global, time-dependent linear analysis. In particular, there are circumstances in which $\tau_{\pi/2}$ is a poor choice of reference time because $\vert\mathcal{M}\vert$ actually peaks somewhat later than this (see \S\ref{sec:Volterra_Long}).

Nevertheless, following the heuristic discussion in \S\ref{sec:Characteristic_Scales}, we expect our disk to always be in the regime \eqref{eqn:small_kernel} if (i) $Q$ is sufficiently large, or (ii) $Q\sim 1$ but we consider sufficiently short wavelengths (so that self-gravity is overwhelmed by velocity dispersion). We justify (i) in Appendix \ref{sec:large_Q}, and we will demonstrate (ii) explicitly in \S\ref{sec:Volterra_short}\footnote{Note that we should \textit{not} expect \eqref{eqn:small_kernel} to be satisfied just by going to very long wavelengths 
 since the shear time \eqref{eqn:tshear} is roughly independent of wavelength.}.
By contrast, given the amplification requirements listed in \S\ref{sec:amplitude_requirements}, many spirals must exist in the regime where self-gravity is very important, leading to  $\mathcal{A}\gtrsim 10^2$ (see equation \eqref{eqn:amplitude_vs_observation}). There will also be cases in-between these two regimes, in which self-gravity is formally strong and yet the system still cannot generate observable spirals.


Thus, the kernel \eqref{eqn:Volterra_Kernel} will be a central object of study in this paper. We now write it in a slightly more explicit form. Without loss of generality, we write our basis functions as
\begin{equation}
    \phi^{(p)}(\br)  = u^q(R)\,\me^{i\ell\varphi},
    \label{eqn:general_basis}
\end{equation}
where we have put $(p)=\ell,q$; thus, $\ell = \pm1, \pm 2, ...$ will label the order of the azimuthal harmonic and $q$ will label the order of the as-yet-unspecified (and generally complex) radial function $u^q(R)$. These have Fourier coefficients
\begin{equation}
   \phi_{\bn}^{(p)}(\bJ)   =   \delta_{n_\varphi}^\ell u^q_{\bn}(\bJ).
   \label{eqn:general_basis_Fourier}
\end{equation}
Moreover, without approximation we can always write
\begin{align}
    u^q(R) = \vert u^q(R)\vert \,\me^{i\int_0^R\md R' w^q(R')},
    \label{eqn:WKB_WLOG}
\end{align}
where $\vert u^q(R) \vert$ and $w^q(R) \equiv \md (\arg u^q)/\md R$ are both real.
Then the wavenumber corresponding to basis element $p$ is $\bk^{(p)} = (k_\varphi^\ell, k_R^q)$, where
\begin{align}
k^\ell_\varphi &= \frac{ \ell }{R_\mathrm{g}},
    \label{eqn:Azimuthal_Wavenumber_q} \\
    k_R^q &=  \frac{-i}{u^q(\Rg)}\frac{\partial u^q(\Rg)}{\p \Rg} 
    = \left(-i \frac{\md \ln \vert u^q\vert}{\md R} + w^q \right)_{R=R_\mathrm{g}}.
    \label{eqn:Radial_Wavenumber_q}
\end{align}
Putting \eqref{eqn:general_basis_Fourier} into \eqref{eqn:Volterra_Kernel}, the Volterra kernel reads 
\begin{align}
    \mathcal{M}^{pp'}(\tau) = \delta_{\ell}^{\ell'}\mathcal{M}^{\ell qq'}(\tau),
    \label{eqn:Volterra_kernel_reads}
    \end{align}
    where
\begin{align}
    \mathcal{M}^{\ell qq'}(\tau) = & -\frac{(2\pi)^2i}{\mathcal{E}} \sum_{n_R}\int \md \bm{J} \, \left( \ell\frac{\p f_0}{\p J_\varphi} +  n_R \frac{\p f_0}{\p J_R} \right)
    \nn
    \\
   & \times 
    [u^{q}_{\ell n_R} (\bm{J})]^*
    u^{q'}_{\ell n_R} (\bm{J}) \me^{ - \mi (\ell \Omega_\varphi + n_R\Omega_R)\tau }.
    \label{eqn:Volterra_Kernel_l}
\end{align} 
This is a rather complicated object, but luckily we can simplify it by using the
results we developed in Paper I. There, we defined the long and short wavelength regimes by $ka_R\sim \epsilon_R$ and $ka_R\sim \epsilon_R^{-1}$ respectively; the intermediate wavelength regime was $ka_R\sim 1$. We then developed asymptotic expressions for potential fluctuations in each regime (and found that these expressions connected smoothly). Since we are now considering the evolution of an entire population of disk stars rather than individual orbits, we will use the rms values $\epsilon, a$ in place of $\epsilon_R$, $a_R$ when defining our asymptotic ordering (see also \S5 of Paper I). The key point is that in each regime the results of Paper I give us explicit expressions for the functions $u_{\ell n_R}(\bJ)$ that enter the kernel \eqref{eqn:Volterra_Kernel_l}. We put these to use in \S\S\ref{sec:Long_Wavelengths}-\ref{sec:Short_Wavelengths} where we investigate spiral structure at long and short wavelengths respectively. First, though, we give a few examples of the behavior of \eqref{eqn:Volterra_Kernel_l} without any wavelength approximations.

\subsubsection{Examples}
\label{sec:Linear_Theory_Example}

As in Paper I, we consider a galaxy with a flat rotation curve with circular speed $V_0$, and for our background DF we take:
\begin{equation}
    f_0(\bm{J}) \propto \me^{-J_\varphi/J_\varphi^\mathrm{s}}\me^{-J_R/\langle J_R\rangle},
\end{equation}
with $J_\varphi^\mathrm{s} = 4R_0 V_0$ and $\langle J_R \rangle = 0.04 R_0 V_0$ independent of $J_\varphi$ (hence $\epsilon \sim 0.1$). The constant $R_0$ has units of length (in a Milky-Way-like context, one can think of $V_0=220$ km\,s$^{-1}$ and $R_0=1$\,kpc). For our radial basis elements, we choose nearly-logarithmic spirals\footnote{In Paper I we employed purely logarithmic spirals. Here, however, we use the modified log-spiral form \eqref{eqn:modified_log_spiral} because it has a well-defined surface density counterpart, allowing us to construct a true biorthogonal basis (equations \eqref{def_basis_1}-\eqref{def_basis_2}).}
(\citealt{Binney2008-ou}, \S{2.6.3}):
\begin{align}
    u^q(R) \propto \,\frac{1}{\sqrt{R}} \, \me^{i\ell \cot\alpha \ln (R/R_0)},
    \label{eqn:modified_log_spiral}
\end{align}
so that $q=(\ell, \alpha)$. From \eqref{eqn:Radial_Wavenumber_q} the radial wavenumber is 
\begin{align}
     k_R^q = \frac{1}{R_\mathrm{g}}\left( \frac{i}{2}+\ell\cot\alpha \right).
     \label{eqn:nearlylogspiral_wavenumber_components}
\end{align}
and so the absolute value of the wavevector $k^{(p)} \equiv \vert \bk^{(p)} \vert$ is 
\begin{align}
    k^{(p)} = \frac{1}{R_\mathrm{g}} \bigg\vert \frac{ 1}{4}+\frac{\ell^2}{\sin^2\alpha} \bigg\vert^{1/2}.
    \label{eqn:nearlyslogspiral_wavenumber_modulus}
\end{align}

\begin{figure*}
    \centering
    \includegraphics[width=0.99\linewidth]{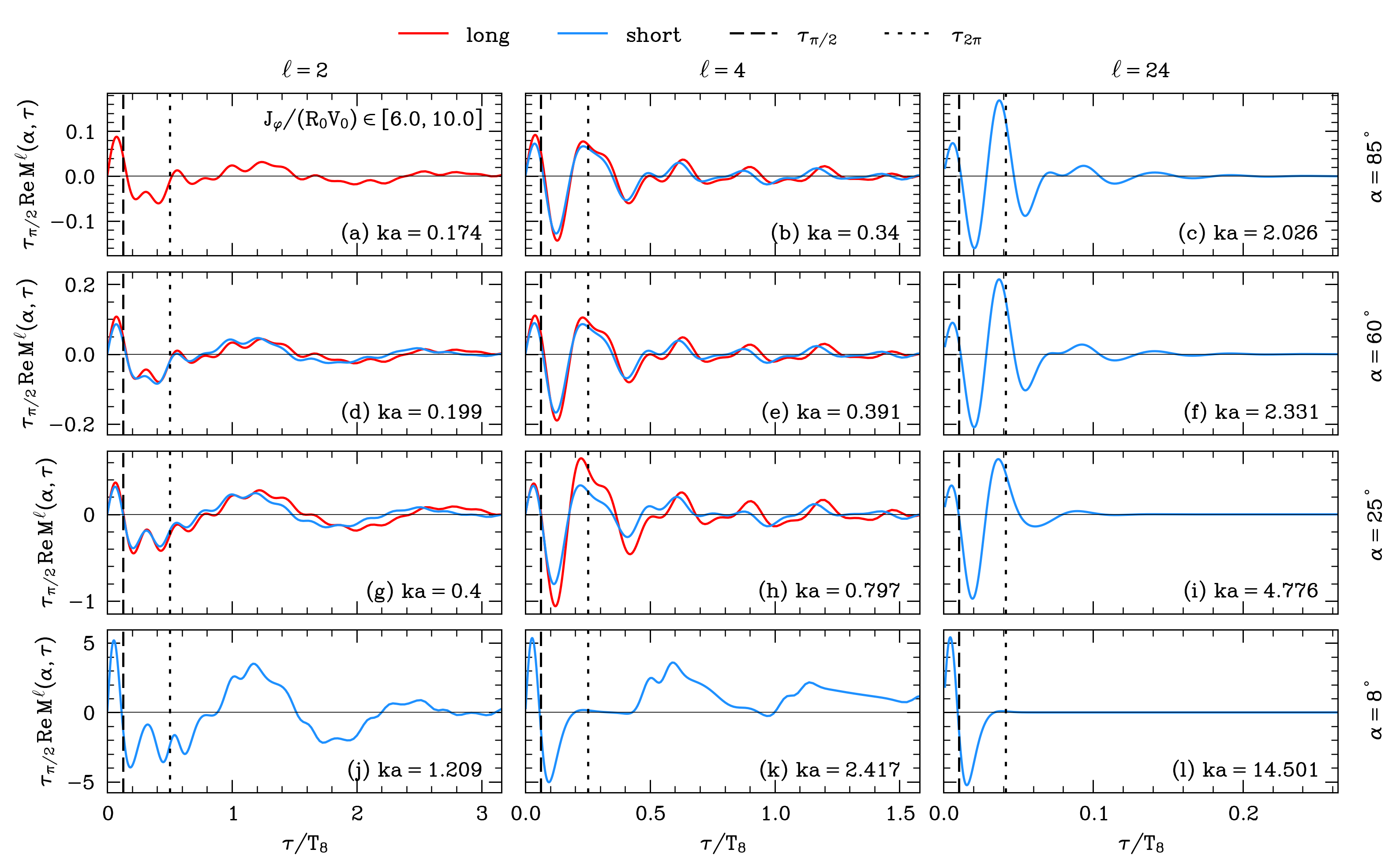}
    \caption{Diagonal element of the Volterra kernel for the basis function \eqref{eqn:modified_log_spiral}, for various combinations of $\ell$ (different columns) and $\alpha$ (different rows).
    Red and blue lines show long/short wavelength analytic approximations respectively. The vertical dotted and dashed black lines show $\tau = \tau_{\pi/2}$ (equation \eqref{eqn:tau_pi_by_2}) and $\tau = \tau_{2\pi} \equiv 4\tau_{\pi/2}$ respectively. Note that each fixed-$\ell$ column has the same horizontal axis, and each fixed-$\alpha$ row has the same vertical axis.}
    \label{fig:Volterra_Kernel_Illustration}
\end{figure*}

In Figure \ref{fig:Volterra_Kernel_Illustration}, we plot some diagonal ($\ell=\ell'$ and $\alpha=\alpha'$) elements of the kernel \eqref{eqn:Volterra_Kernel_l}, which we call $\mathcal{M}^\ell(\alpha, \tau)$, for this choice of basis function. To perform this calculation, we
restricted the $J_\varphi$ range over which we integrate to $J_\varphi/(R_0V_0)\in(6, 10)$; the typical dynamical time is then $t_\mathrm{dyn} \sim T_8\equiv 2\pi (8R_0)/V_0$.
The mass $M$ contained in stars in this $J_\varphi$ range enters the kernel via an 
overall prefactor that we call
\begin{equation}
    Y \equiv \frac{M}{M_\mathrm{enc}}, \,\,\,\,\,\,\,\,\,\,\,\,\,M_\mathrm{enc} = \frac{8R_0V_0^2}{G}.
    \label{eqn:active_fraction}
\end{equation}
This is the \textit{active fraction}, the ratio of $M$ to the total mass $M_\mathrm{enc}$ enclosed within $8R_0$, as implied by the rotation curve. We choose $Y=0.1$, 
but choosing a different value of $Y$ would just correspond to amplifying/shrinking the curves in Figure \ref{fig:Volterra_Kernel_Illustration} by the appropriate factor.

In fact, since a brute-force numerical calculation of \eqref{eqn:Volterra_Kernel_l} incurs considerable computational expense, we do not pursue it here. Instead, the red and blue lines in Figure \ref{fig:Volterra_Kernel_Illustration} correspond to long- and short-wavelength approximations to the kernel \eqref{eqn:Volterra_Kernel_l}, valid for $ka\sim \epsilon$ and $ka\sim \epsilon^{-1}$ respectively. We will discuss these asymptotic results in detail in \S\S\ref{sec:Long_Wavelengths}-\ref{sec:Short_Wavelengths}; for now we note that they rely only upon the approximations for $u_{\bn}^q(\bJ)$ that we developed, and verified numerically, in Paper I\footnote{Strictly speaking, in Paper I we verified numerically that our approximations are accurate to $\mathcal{O}(\epsilon^2)$ for \textit{pure} logarithmic spirals, not the \textit{nearly}-logarithmic spirals we use here, but we have performed the same verification for the nearly-logarithmic spirals also.}. Moreover,in Paper I we found that these results connected smoothly at intermediate wavelengths, $ka\sim 1$. As such, in some panels of Figure \ref{fig:Volterra_Kernel_Illustration} we show both long and short wavelength results (the label in each panel gives the corresponding value of $ka$). We find that 
the smooth transition occurs somewhere in the range $ka \in (0.3, 0.5)$, although this is dependent on the precise parameters we choose and would change somewhat if we used different basis functions.

The twelve panels of Figure \ref{fig:Volterra_Kernel_Illustration} show $\tau_{\pi/2} \mathrm{Re}\,\mathcal{M}^{\ell}(\alpha, \tau)$ for different values of $\ell$ (different columns) and $\alpha$ (different rows). With vertical black dashed and dotted lines we indicate two important $\tau$ values, namely $\tau_{\pi/2}$ (equation \eqref{eqn:tau_pi_by_2}) and $\tau_{2\pi} = T_8/m = 4\tau_{\pi/2}$. Note that the axes are not the same across all panels, although each row (fixed $\alpha$) has the same vertical axis, and each column (fixed $\ell$) has the same horizontal axis.
In all panels of Figure \ref{fig:Volterra_Kernel_Illustration} we see that the kernel undergoes the expected oscillations and overall envelope decay. We have checked that in each case the envelope decays roughly $\propto \tau^{-s}$, where $s\in (1,2)$; the value $s=1$ is what we would expect from integrating the rapidly-oscillating integrand in \eqref{eqn:Volterra_Kernel_l} over action at large $\tau$, while $s=2$ is what we get when the Dehnen drift to the azimuthal frequency is dominant (see below).  The kernel tends to peak in absolute values around $\tau \sim \tau_{\pi/2}$, and thereafter oscillates. Roughly speaking, the smaller is $ka$, the fewer harmonics are involved in this oscillation, whereas for large $ka$ the pattern can be very complicated. The main thing that changes when we vary $\alpha$ is the peak \textit{amplitude} of the kernel --- see \S\ref{sec:Volterra_short} for further discussion.
\section{Long wavelengths}
\label{sec:Long_Wavelengths}

In this section, we consider spirals with long wavelengths, $k\sim R_\mathrm{g}^{-1}$ (i.e., $ka\sim\epsilon$), which necessarily requires $m\sim 1$. To begin, we show how to simplify the Volterra kernel in this regime (\S\ref{sec:Volterra_Long}). We then discuss two contrasting cases: first when the spiral's self-gravity is negligible (\S\ref{sec:Long_Wavelengths_Weak_Self_Gravity}), and second when the disk is globally unstable to spiral modes (\S\ref{sec:Global_Instabilities}).

\subsection{Volterra kernel}
\label{sec:Volterra_Long}

At long wavelengths, we know from Paper I that the only contributions to the basis functions $\phi^{(p)}_{\bm{n}} (\bm{J})$ will be from $n_R=0,\pm 1$. In Appendix \ref{sec:Long_wavelength_Kernel} we show that, for a Schwarzschild-type DF
\begin{equation}
    f_0(\bm{J}) = g_0(J_\varphi) \me^{-J_R/\langle J_R\rangle},
    \label{eqn:Schwarzchild}
\end{equation}
where $\langle J_R \rangle$ may depend on $J_\varphi$, the Volterra kernel \eqref{eqn:Volterra_Kernel_l} is equal to
\begin{align}
    \mathcal{M}^{\ell qq'}(\tau) &= 
    \frac{1}{\mathcal{E}} 
    \int \md J_\varphi [u^{q}(\Rg)]^*u^{q'}(\Rg) \bigg[ G^{\ell}_{0} \me^{-i\ell \Omega\tau} 
    \nn
    \\ & +
    G^{\ell q q'}_{+} \me^{-i(\ell\Omega+\kappa)\tau} 
    +
    G^{\ell q q'}_{-}
    \me^{-i(\ell\Omega-\kappa)\tau} 
    \bigg],
    \label{eqn:Volterra_Kernel_Long}
\end{align} 
where the auxiliary functions $G^{\ell}_0(J_\varphi, \tau)$ and $G^{\ell qq'}_\pm(J_\varphi, \tau)$, which have the same units as $g_0$, are given in equations \eqref{eqn:aux_0}-\eqref{eqn:aux_pm}. Importantly, the only $\tau$-dependence in these functions arises through various factors of 
\begin{align}
    \zeta(\Rg, \tau) \equiv 1+i\ell \langle \Omega_\mathrm{D} \rangle \tau,
    \label{eqn:Dehnen_phase}
\end{align}
where
\begin{equation}
   \langle  \Omega_\mathrm{D} \rangle \equiv \langle J_R \rangle \frac{\md \kappa}{\md J_\varphi} \sim \epsilon^2 \Omega,
    \label{eqn:Dehnen}
\end{equation}
is the {mean} Dehnen drift frequency of the population of stars with angular momentum $J_\varphi$ (see \S3.2 of Paper I). For small enough $\tau$ we can set $\zeta \simeq 1$, in which case the only time dependence of the Volterra kernel is that given explicitly in \eqref{eqn:Volterra_Kernel_Long}. The red lines in Figure \ref{fig:Volterra_Kernel_Illustration} show the long-wavelength approximation \eqref{eqn:Volterra_Kernel_Long} to the full kernel \eqref{eqn:Volterra_Kernel_l} for the specific examples discussed in \S\ref{sec:Linear_Theory_Example}.

Following the discussion of \S\ref{sec:Characteristic_Scales}, we might expect self-gravity to be unimportant on timescales $\gtrsim t_\mathrm{shear} \sim \Omega^{-1} \sim t_\mathrm{dyn}$ (equation \eqref{eqn:tshear}) on the very largest scales in our disk, since it will eventually be overwhelmed by shear. Mathematically, this follows from \eqref{eqn:Volterra_Kernel_Long}: a very long wavelength basis function $u^q$ will be nearly constant on the scale $\sim \Rg$, while the factors $\me^{-i(\ell \Omega + n_R\kappa)\tau}$ will usually be rapidly oscillating functions of action for $\tau \gtrsim \Omega^{-1}$. The exception will be for any combination of frequencies $\omega = \ell \Omega + n_R\kappa$ that does \textit{not} vary rapidly with $J_\varphi$. By far the most important combinations are multiples of the `Lindblad-Kalnajs' frequency
\begin{equation}
   \Omega^\mathrm{LK} \equiv   \Omega - \frac{\kappa}{2};
   \label{eqn:Lindblad_Kalnajs}
\end{equation}
this is sometimes also referred to as $\Omega^\mathrm{ILR}$  --- for `inner Lindblad resonance' --- in the literature. This frequency is special because in many galactic disk models it is $\lesssim \Omega/3$ and nearly constant over many kpc in guiding radius. Fluctuations that rotate at some multiple of $\Omega^\mathrm{LK}$ can therefore resist shear for a long time, and play a crucial role in long-wavelength spiral structure both with and without self-gravity, as we will see.

\begin{figure}
    \centering
    \includegraphics[width=0.99\linewidth]{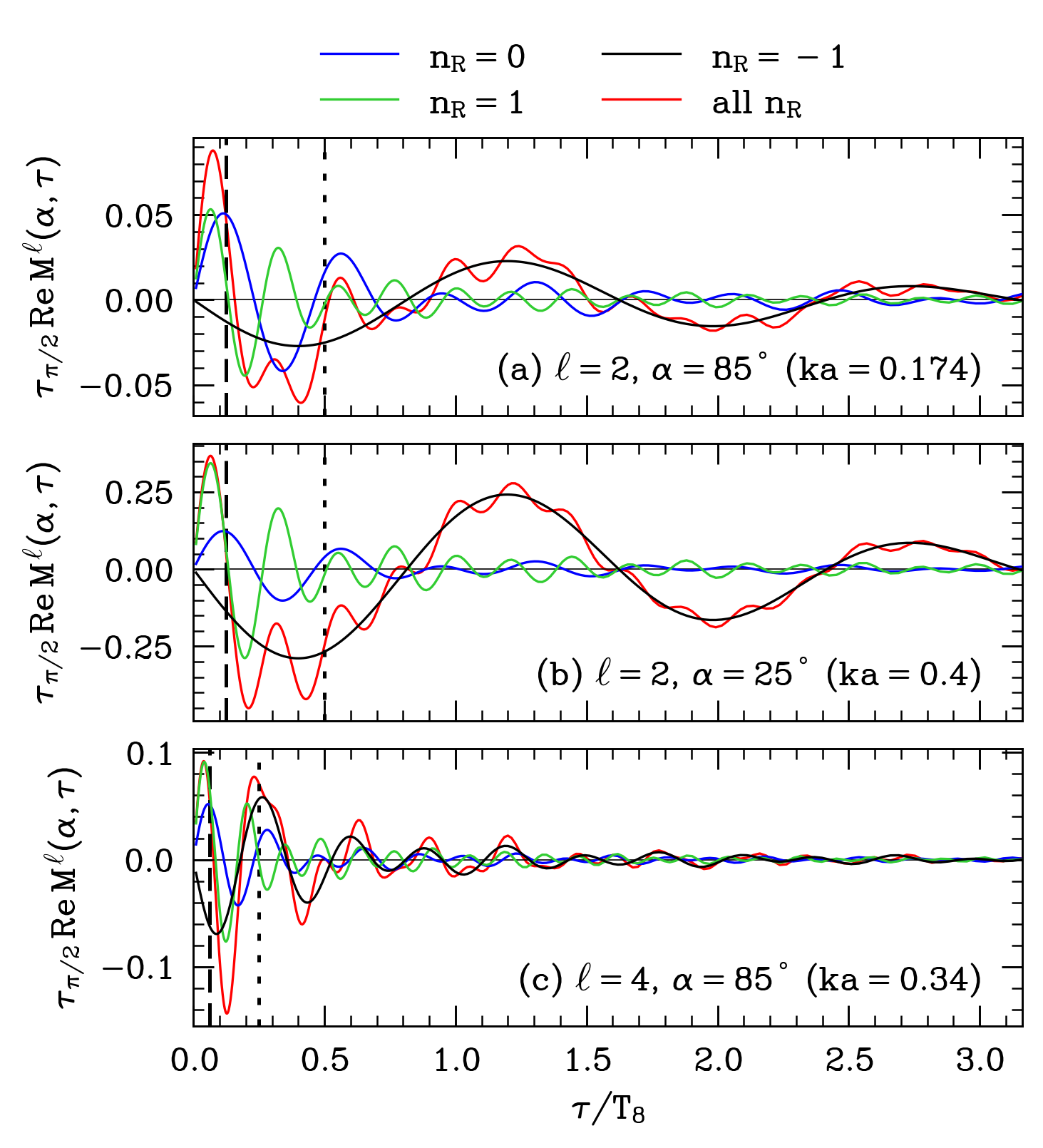}
    \caption{Long wavelength kernel \eqref{eqn:Volterra_Kernel_Long} for the same log-spiral basis as in \S\ref{sec:Linear_Theory_Example}. The red lines show the full result \eqref{eqn:Volterra_Kernel_Long}
    while the other colored lines show the contributions from individual $n_R$ values (corresponding to terms $\propto \me^{-i(\ell\Omega+n_R\kappa)\tau}$ in \eqref{eqn:Volterra_Kernel_Long}).}
    \label{fig:Volterra_Kernel_Illustration_nR_Dependence}
\end{figure}

As an example, in Figure \ref{fig:Volterra_Kernel_Illustration_nR_Dependence} we re-plot some of the red long-wavelength curves from Figure \ref{fig:Volterra_Kernel_Illustration}. Precisely, in panels (a) and (b) we set $\ell=2$, but consider two very different pitch angles $\alpha$, namely (a) $85^\circ$ and (b) $25^\circ$; panel (c) is again for $\alpha=85^\circ$ but now $\ell=4$.  In addition, with blue, black and green lines we show the individual contributions to \eqref{eqn:Volterra_Kernel_Long} coming from particular $n_R$ values (corresponding to terms $\propto \me^{-i(\ell\Omega+n_R\kappa)\tau}$). We see that in both panels (a) and (b) the dominant contribution for $\tau \gtrsim \tau_{2\pi}$ is from $n_R=-1$, i.e., the term $\propto \me^{-2i\Omega^\mathrm{LK}\tau}$ in \eqref{eqn:Volterra_Kernel_Long}. This dominance is more apparent in panel (b) than in panel (a). Physically, this is because in panel (a) the spiral is so open that, at a fixed azimuthal phase, the potential felt by a given star does not depend much on its radial phase.
Though we do not show it here, at a fixed $\alpha$ the dominance of $n_R=-1$ becomes ever stronger if we extend the $J_\varphi$ range over which we integrate. By contrast, in panel (c) there is no possibility of oscillations at a multiple of the frequency \eqref{eqn:Lindblad_Kalnajs}, so the $n_R=-1$ contribution no longer overwhelms the others.

\subsection{Weak self-gravity}
\label{sec:Long_Wavelengths_Weak_Self_Gravity}

Let us now suppose self-gravity is negligible, meaning equation \eqref{eqn:small_kernel} is satisfied for all $\tau$. Physically this is likely to be the case if we consider a disk that is sufficiently `hot' (see Appendix \ref{sec:large_Q}).
In this regime, stars act like test particles: they orbit in the mean field potential and respond to any external forcing $\delta \phi^\mathrm{ext}$, but do not communicate with one another. Because of this, we need not use the Volterra equation \eqref{eqn:Volterra_Equation}. Instead we can calculate the DF fluctuation directly from \eqref{eq:linear_Vlasov_formal_solution} with $\delta \phi^\mathrm{tot} = \delta \phi^\mathrm{ext}$.

Without self-gravitating amplification, the only way to get observable spiral structure (as we argued in \S\ref{sec:amplitude_requirements}) is to force the disk `externally', for instance with a satellite galaxy interaction (e.g., \citealt{dobbs2010simulations}). Here we will consider only the simplest case of an initially unperturbed disk subjected to an impulsive kick at $t=0$:
\begin{equation}
    \delta \phi^\mathrm{ext}(\br,t)
    =
    T\delta(t)U(\br),
    \end{equation}
where $T$ has units of time. Plugging this into \eqref{eq:linear_Vlasov_formal_solution}, we 
find that for $t>0$ our `spiral' simply phase mixes in the mean potential:
\begin{equation}
     \delta f_{\bn}(\bJ,t) = i\bn\cdot\frac{\p f_0}{\p \bJ}TU_{\bn}(\bJ)\me^{-i\bn\cdot\bOm t}.
     \label{eqn:impulsive_solution}
\end{equation}
Let us reconstruct the corresponding density fluctuation $\delta \Sigma$ from its basis expansion \eqref{eqn:DeltaSigma_basis}, by plugging \eqref{eqn:impulsive_solution} into \eqref{eqn:B_formal} to get the coefficients $B^p$. Expanding the forcing function 
\begin{equation}
    U(\br) = \sum_{p}C^{p}\phi^{(p)}(\br),
\end{equation}
for some dimensionless coefficients $C^{p}$, we find
\begin{align}
    B^p(t) =  T \sum_{p'}  \mathcal{M}^{pp'}(t)C^{p'},
    \label{eqn:impulsive_master}
\end{align}
where we identified the Volterra kernel $\mathcal{M}^{pp'}$ from \eqref{eqn:Volterra_Kernel}. Thus, even though we are considering a problem without self-gravity, the Volterra kernel \eqref{eqn:Volterra_Kernel}
rears its head anyway. This is advantageous, since it means we can use the results of \S\ref{sec:Volterra_Long} to understand this problem, too.

We now make some simplifying assumptions. First, since a massive perturber's scale radius will typically be $\gtrsim a$, most of the power in $\delta\phi^\mathrm{ext}$ should lie at long and intermediate wavelengths, so we will use the approximate long-wavelength kernel \eqref{eqn:Volterra_Kernel_Long}. Second, given the discussion of \S\ref{sec:Volterra_Long}, we expect the dominant terms in \eqref{eqn:Volterra_Kernel_Long} to be those $\propto \me^{-2i\Omega^\mathrm{LK}\tau}$, so we drop all other terms. Third, we suppose that the disk was smooth before the encounter so we can drop all gradients of $g_0$ and $\langle J_R \rangle$ in the auxiliary functions $G_\pm$ (see equation \eqref{eqn:aux_pm}). Fourth, we make the (rather crude) assumption that $\Omega^\mathrm{LK}$ (equation \eqref{eqn:Lindblad_Kalnajs}) and $\langle J_R \rangle$ do not vary over the $J_\varphi$ range of interest, so we can take the terms involving them out of the $J_\varphi$ integral in \eqref{eqn:Volterra_Kernel_Long}.

Putting these assumptions together, plugging the resulting kernel into \eqref{eqn:impulsive_master}, and inserting this into 
\eqref{eqn:DeltaSigma_basis}, we find
  \begin{align}
&\delta\Sigma(\br, t) \simeq  \sum_q \left[  \sum_{\pm}
\frac{\me^{\pm 2i(\varphi-\Omega^\mathrm{LK}t)}}{(1\pm 2i\langle \Omega_\mathrm{D}\rangle t)^2}b_\pm^q \right] \Sigma^q(R) .
\label{eqn:explicit_deltaSigma_impulsive}
\end{align}
where the time-independent, dimensionless quantities $b_\pm^q$ are straightforward to derive but unenlightening. Equation \eqref{eqn:explicit_deltaSigma_impulsive} corresponds to a two-armed spiral rotating {nearly} rigidly with pattern speed \eqref{eqn:Lindblad_Kalnajs}. Such spirals appear naturally in systems that receive large-scale impulsive kicks, such as that by \cite{bland2021galactic} of a Sagittarius-like perturber encountering a Milky-Way-like galaxy. We emphasize, though, that the rotation of these spirals is only \textit{nearly} rigid, for two reasons. First, the frequency $\Omega^\mathrm{LK}$ is not truly constant over the extent of the disk, so we have over-simplified things by treating it as fixed. Second, there is additional time-dependence in the denominator of \eqref{eqn:explicit_deltaSigma_impulsive} owing to the Dehnen drift of the azimuthal frequency. At late times $t \gtrsim (2\langle \Omega_\mathrm{D}\rangle )^{-1}$, this term dominates, and the spiral amplitude decays like $\vert \delta \Sigma \vert \propto t^{-2}$.

\subsubsection{Why are so many spirals two-armed?}
\label{sec:Why_Two_Armed}


The idealized, two-armed, non-self-gravitating spirals that we just uncovered, rotating prograde with pattern speed $\Omega^\mathrm{LK}$, are called two-armed \textit{Lindblad-Kalnajs kinematic density waves}. (See Appendix \ref{sec:density_waves} for a synopsis of the --- often confusing --- terminology surrounding density waves).
More generally, one could set up $m$-armed versions of these waves rotating with pattern speed $\Omega^\mathrm{LK}$. In practice, though, this is only achievable for even $m$, because the response $\delta f_{\ell n_R}(\bJ,t)$ in the absence of self-gravity most often has time-dependence $\propto \me^{-i\ell (\Omega+n_R\kappa /\ell)t}$ --- see equation \eqref{eqn:impulsive_solution} --- and the combination $\Omega+n_R\kappa /\ell$ can only equal $\Omega^\mathrm{LK}$ if $m=\vert \ell\vert$ is even. In fact, at long/intermediate wavelengths, these spirals are naturally two armed, rather than four-armed, six-armed, or whatever. This is because most power is invested in $n_R = 0,\pm 1$, so the only combinations $\Omega+n_R\kappa /\ell$ that can equal $\Omega^\mathrm{LK}$ are $(\ell, n_R)=\pm(2,-1)$.\footnote{This is different from the traditional argument in terms of uniformly-precessing nested ellipses \citep{Binney2008-ou}. While the traditional argument explains why Lindblad-Kalnajs patterns are possible (and long-lived), it does not tell us why $m=2$ should be preferred, since one could orient the ellipses to produce a pattern that preferred any even $m$.}


\subsection{Global instability}
\label{sec:Global_Instabilities}

We now consider those long wavelength spirals for which self-gravity {cannot} be ignored. Although a general linear theory in this regime would involve solving the Volterra equation \eqref{eqn:Volterra_Equation} with the long-wavelength kernel \eqref{eqn:Volterra_Kernel_Long}, we will focus on a simpler problem, namely the generation of global instabilities.

A spiral instability is a perturbation whose only time dependence, apart from an overall rigid rotation, is an exponential growth in amplitude.  Suppose such a perturbation exists, and let us choose the first of our basis elements $\phi^{(p)}(\br)$ to have its spatial form. Then we expect the corresponding coefficient $B^p$ (see equation \eqref{eqn:phi_basis}) to satisfy $B^p(t) = B^p(0)\me^{-i\omega t}$, where Im\,$\omega>0$ is the growth rate of the instability, and Re\,$\omega$ is the rotation frequency. At late times this should dominate the solution to the Volterra equation \eqref{eqn:Volterra_Equation}, so we can drop the $B_\mathrm{kin}$ and $B_\mathrm{ext}$ terms in that equation and take the upper limit of the $t'$ integral to infinity, resulting in
\begin{align}
    1 - \widehat{M}(\omega) = 0,
    \label{eqn:Landau_Dispersion_Relation_Long}
\end{align}
where $\widehat{M}$ is the first (diagonal) element of the Landau response matrix $\widehat{\mathcal{M}}^{pp'}(\omega)$ (equation \eqref{Fourier_M}). In other words, our instability is a Landau mode (equation \eqref{eqn:Landau_Dispersion_Relation}) with Im\,$\omega >0$.

Plugging  \eqref{eqn:Volterra_Kernel_Long} into \eqref{eqn:Landau_Dispersion_Relation_Long} and dropping all superscripts, the long-wavelength `dispersion relation' is
\begin{align}
    1 = 
    \frac{1}{\mathcal{E}} &
    \int  \md J_\varphi \vert u(\Rg) \vert^2 \bigg[ \widehat{G}_{0}(\omega -\ell \Omega) 
    \nn
    \\ & +
    \widehat{G}_{+}[\omega -(\ell \Omega+\kappa)]
    +
    \widehat{G}_{-}[\omega -(\ell \Omega-\kappa)]
    \bigg].
    \label{eqn:Dispersion_Relation_Long}
\end{align} 
where the hats denote Laplace transforms as in \eqref{eqn:Laplace}.
Now,  $\widehat{G}_0$ and $\widehat{G}_\pm$ (see equations \eqref{eqn:aux_0}-\eqref{eqn:aux_pm}) involve the Laplace transforms of $\zeta^{-n}(\tau)$ where $\zeta$ is the Dehnen factor from equation \eqref{eqn:Dehnen_phase} and $n=1,2,3$. These can be calculated analytically but the resulting expressions are very cumbersome, so for simplicity we will assume that the Dehnen drift can be ignored throughout the rest of this section (i.e., we fix $\zeta=1$). Then we find 
\begin{align}
    &1 = 
    \frac{(2\pi)^2\ell}{\mathcal{E}} 
    \int  \md J_\varphi \vert u(\Rg) \vert^2 \nn \bigg[\frac{1}{\omega -\ell \Omega} \frac{\md (g_0 \langle J_R\rangle)}{\md J_\varphi}  \, + \nn
    \\  & \,\,\,\,\,\,\,\sum_{\pm}\frac{\xi_{\pm}(R_\mathrm{g},a)}{\omega -(\ell \Omega\pm\kappa)}\left(\langle J_R\rangle \frac{\md g_0}{\md J_\varphi} + 2 g_0 \frac{\md \langle J_R \rangle}{\md J_\varphi} \mp \frac{g_0}{\ell}\right)
    \bigg],
    \label{eqn:Dispersion_noDehnen}
\end{align} 
where $\xi_{\pm}$ is defined in \eqref{eqn:dimensionless_functions} (and we set $q=q'$). 

In the limit Im\,$\omega\to0^+$, the right hand side of \eqref{eqn:Dispersion_noDehnen} is a sum of terms associated with corotation and Lindblad resonances, though more generally these resonances are broadened in frequency space with a  width $\sim \vert \mathrm{Im}\, \omega\vert$. The sign of the corotation contribution depends on whether $g_0\langle J_R \rangle$ is increasing or decreasing around the resonance.

Let us now consider the `dispersion relation' \eqref{eqn:Dispersion_noDehnen} in a few different limits.

\subsubsection{Smooth DF}
\label{sec:smooth_DF}


If there are no sharp features in the DF we expect 
\begin{align}
    \langle J_R\rangle  \bigg\vert  \frac{\md g_0}{\md J_\varphi} \bigg\vert \sim g_0 \bigg\vert  \frac{\md \langle J_R \rangle}{\md J_\varphi}\bigg\vert &\sim 
    \frac{\langle J_R\rangle g_0}{J_\varphi} 
    \sim  \epsilon^2 g_0,
\end{align}
meaning that the `gradient' terms are negligible compared to $g_0/\ell$ on the second line in \eqref{eqn:Dispersion_noDehnen}. 
Thus for a smooth DF we get
\begin{align}
    &1 = 
    \frac{(2\pi)^2\ell}{\mathcal{E}} 
    \int  \md J_\varphi \vert u(\Rg) \vert^2 \nn \bigg[\frac{1}{\omega -\ell \Omega} \frac{\md (g_0 \langle J_R\rangle)}{\md J_\varphi}  \, + \nn
    \\  & \,\,\,\,\,\,\,\sum_{\pm}\frac{\mp \xi_{\pm}(R_\mathrm{g},a)}{\omega -(\ell \Omega\pm\kappa)}  \frac{g_0}{\ell} 
    \bigg],
    \label{eqn:Dispersion_noDehnen_Smooth}
\end{align} 
We do not throw away the first (`corotation') term, because the prefactor $\xi_{\pm}$ in front of the Lindblad terms is typically $\sim \epsilon^2$ (see equation \eqref{eqn:dimensionless_functions}), meaning all terms in the square bracket are formally comparable. 

There are, however, circumstances in which one term in \eqref{eqn:Dispersion_noDehnen_Smooth} dominates. For instance, $g_0$ can be much larger near the inner Lindblad resonance than either of the other resonances. If we keep only the inner Lindblad term (the lower sign on the second line of \eqref{eqn:Dispersion_noDehnen_Smooth} in the case $\ell = +m>0$) then we recover a dispersion relation for bar instability --- see, e.g., \S12.7 of \cite{palmer1994stability}.

Another interesting limit of \eqref{eqn:Dispersion_noDehnen_Smooth} is that of a very cold disk, $\langle J_R \rangle \to 0$. Naively, one might expect the whole right hand side of \eqref{eqn:Dispersion_noDehnen_Smooth} should then vanish, meaning no instabilities are possible. But this is incorrect, because the combination  $\langle J_R \rangle g_0$ is independent of $\langle J_R \rangle$ for a fixed surface density profile (Appendix \ref{sec:large_Q}), and so is finite even as $\langle J_R \rangle \to 0$. This allows for the possibility that a disk of stone-cold orbits possesses long-wavelength nonaxisymmetric instabilities, in addition to the well-known short-wavelength axisymmetric instabilities that exist whenever $Q<1$.\footnote{These follow from the LSK dispersion relation (\S\ref{sec:LSK}).}


\subsubsection{Sharp DF}
\label{sec:Long_Wavelength_Instability_Sharp}

Suppose instead that $g_0$ has a sharp feature like a groove \citep{sellwood1991spiral,De_Rijcke2019-uo}, centered on $J_\varphi = J_*$,  corresponding to guiding radius $R_\mathrm{g}(J_*)\equiv R_*$,  with characteristic width $\Delta J \ll J_*$. Then in the corotation term in \eqref{eqn:Dispersion_noDehnen}, we can write  $\md (g_0\langle J_R\rangle)/\md J_\varphi \simeq \langle J_R\rangle\, \md g_0/\md J_\varphi$ up to relative corrections   $\mathcal{O}(\epsilon^2)$ provided $\Delta J/J_* \lesssim \epsilon^2$, i.e., $\Delta J \lesssim \langle J_R \rangle$. The term involving the gradient of $g_0$ may also dominate the Lindblad portion of \eqref{eqn:Dispersion_noDehnen}, but since $\vert \xi_\pm(R_\mathrm{g}, a)\vert  \sim \epsilon^2$  these will tend to be subdominant anyway compared to corotation. With these simplifications,
\begin{align}
    &1 \simeq
    \frac{(2\pi)^2\ell}{\mathcal{E}} 
    \int  \md J_\varphi \vert u(\Rg) \vert^2  \frac{\langle J_R\rangle }{\omega -\ell \Omega} \frac{\md g_0}{\md J_\varphi}.
    \label{eqn:Dispersion_noDehnen_CROnly}
\end{align} 
Now, expanding
\begin{equation}
    J_\varphi = J_* + j, \,\,\,\,\, R_\mathrm{g} = R_* + r,
    \label{eqn:expanding_R_and_J}
\end{equation}
with $\vert j \vert \ll J_*$ and $\vert r\vert \ll R_*$, and assuming the groove is narrow enough that $u^q$ hardly varies across it, we get 
\begin{align}
    &1 \simeq
    \frac{(2\pi)^2\ell}{\mathcal{E}} \vert u(R_*) \vert^2\langle J_R\rangle
    \int_{-\infty}^\infty  \md j \,\frac{\md g_0/\md j}{\omega -\ell \Omega_* - \ell \Omega'_* j} ,
    \label{eqn:Dispersion_noDehnen_CROnly_expanded}
\end{align} 
where $\Omega_* = \Omega(J_*)$ and $\Omega'_* = [\md\Omega/\md J_\varphi]_{J=J_*}$.

Equation \eqref{eqn:Dispersion_noDehnen_CROnly_expanded} is essentially the classic \cite{sellwood1991spiral} dispersion relation for the groove instability. However, our derivation is slightly more general than theirs, because they had to invoke softening (on a scale much larger than the groove) in order to justify truncating their potential expansion. Of course real galaxies do not have softened gravity, though one might argue that such softening would be an effective outcome of finite velocity dispersion (see \S\ref{sec:Volterra_short}). By contrast, in deriving \eqref{eqn:Dispersion_noDehnen_CROnly_expanded} we have simply searched for an unstable Landau mode with long wavelength and not invoked any softening at all. 
 
We mention that the symmetry $g_0\leftrightarrow \langle J_R \rangle$ in the corotation term in \eqref{eqn:Dispersion_noDehnen} suggests there should also be analogous instabilities associated with grooves in the $\langle J_R \rangle$ profile that share many of the same features as the standard groove instability. Investigation of such instabilities is an interesting avenue for future work.

The above analysis is somewhat oversimplified, especially if one is interested not in grooves (localized underdensities in $g_0$) but ridges (localized overdensities) or in other sharp features like disk breaks \citep{fiteni2024role}. In fact, one can show that \eqref{eqn:Dispersion_noDehnen_CROnly_expanded} on its own implies stability in both of these cases, contrary to experiment \citep{sellwood1991spiral}. Thus, one must work with a less approximate dispersion relation in order to recover the ridge/break instabilities. The most obvious generalization is to include some Lindblad contributions from equation \eqref{eqn:Dispersion_noDehnen}. A more subtle consideration is that sharp features in the DF tends to introduce a local kink in the profile of radial frequencies $\kappa(J_\varphi)$ --- see, e.g., Figure 3 of \cite{fiteni2024role} --- meaning the Dehnen drift \eqref{eqn:Dehnen} might be (locally) very large, and we should not use equation \eqref{eqn:Dispersion_noDehnen} at all but rather the more general \eqref{eqn:Dispersion_Relation_Long}. Further exploration of these ideas is beyond the scope of this paper.

\subsubsection{Maximum growth rate}
\label{sec:max_growth_rate}

An interesting corollary of the above analysis is that one can deduce a rough maximum growth rate for spiral instabilities. Technically we have to specify that we are considering only disks that have $Q$ in the vicinity of unity, i.e., we have to exclude pathological cases like stone-cold disks in which self-gravitating collapse happens on an arbitrarily short time (see \S6.2.3 of \citealt{Binney2008-ou}). With this caveat, we will find that no instability ever grows faster than the typical orbital frequency.

To derive this result, recall that a general instability is a solution to \eqref{eqn:Landau_Dispersion_Relation} with $\omega = \omega_\mathrm{R}+i\omega_\mathrm{I}$, for some growth rate $\omega_\mathrm{I}>0$. By maximizing $\omega_\mathrm{I}$ we are making the denominator in \eqref{Fourier_M} rather large, but to solve \eqref{eqn:Landau_Dispersion_Relation} we must still balance the whole right hand side of \eqref{Fourier_M} to unity. A little thought reveals that the maximum $\omega_\mathrm{I}$ will arise when we have a single dominant $\bn$ on the right hand side of \eqref{eqn:Landau_Dispersion_Relation} \textit{and} there is a sharp feature in the DF at precisely the action space location where $\omega_\mathrm{R}\simeq \bn\cdot\bOm$ --- basically, we are dealing with something like the groove instability, with dispersion relation \eqref{eqn:Dispersion_noDehnen_CROnly_expanded}. 

Let us therefore take \eqref{eqn:Dispersion_noDehnen_CROnly_expanded} and ask what is the maximum allowable $\omega_\mathrm{I}$. We integrate by parts with respect to $j$ on the right hand side, and find that the maximum possible value of $\omega_\mathrm{I}$ occurs when $\omega_\mathrm{R}-\ell\Omega_* = 0$. Employing the basic scaling $\mathcal{E}\sim G^{-1} kR^2 \vert u\vert^2$ which follows from equations \eqref{def_basis_1}-\eqref{def_basis_2}, and using the fact that the mass in the groove region is $\Delta M = (2\pi)^2\langle J_R\rangle\int \md j\,g_0$, we can estimate
\begin{equation}
    1 \sim \bigg\vert \frac{\ell^2G \Omega_*' \,\Delta M}{kR_*^2 \omega_\mathrm{I, max}^2} \bigg\vert 
\end{equation}
Now we use the general result
\begin{equation}
    \frac{\md \Omega}{\md J_\varphi} = \frac{A}{BR^2},
\end{equation}
where
\begin{align}
    A \equiv - \frac{1}{2} \frac{\md \Omega}{\md \ln R}, \,\,\,\,\,\,\,\,\,\, B=A-\Omega
\label{eqn:Oort_Constants}
\end{align}
are the Oort constants. Then
\begin{equation}
    \omega_\mathrm{I, max}^2 \sim \frac{\ell^2}{kR} \times \frac{G \,\Delta M}{R^3} \times \bigg\vert \frac{A}{B} \bigg\vert
\end{equation}
and everything is to be evaluated at $R=R_*$. 
An upper bound to the right hand side occurs if $\Delta M$ is comparable to the mass of the system, in which case $G\,\Delta M/R^3 \sim \Omega^2$ and we have 
\begin{equation}
    \omega_\mathrm{I,\mathrm{max}} \sim   \frac{m}{\sqrt{kR}}  \times \Omega \times \bigg\vert \frac{A}{B} \bigg\vert^{1/2} \sim \Omega.
\end{equation}
where for the last estimate we assumed $m\sim kR\sim \vert A/B\vert\sim 1 $.
A similar result was found in the fluid context by \cite{balbus1992oort} (and might have been anticipated from \S\ref{sec:Characteristic_Scales}).
\section{Short wavelengths}
\label{sec:Short_Wavelengths}

Now, we turn to spiral fluctuations at short wavelengths, $k\sim R_\mathrm{g}a^{-2}$ (i.e., $ka\sim\epsilon^{-1}$). To begin we simplify the Volterra kernel in this regime, and discuss the role of self-gravity on the smallest scales (\S\ref{sec:Volterra_short}). Then we consider two further limiting cases. First, we consider the limit of short azimuthal wavelengths (large $m$, see \S\ref{sec:Small_Azimuthal}), where we recover (and generalize) the classic results of the shearing sheet. Then, we look at the tight-winding limit ($m\sim 1$ and small radial wavelengths, see \S\ref{sec:Small_Radial}) and recover the LSK dispersion relation. 

\subsection{Volterra kernel}
\label{sec:Volterra_short}

At short wavelengths we take the basis functions to be a sum of sinusoidal waves, i.e., we take $\vert u^q\vert$ and $w^q$ to be constants (or rather, very slowly varying functions of $R$) in equation \eqref{eqn:WKB_WLOG}. In particular $k_R^q = w^q(R_\mathrm{g})$ (see equation \eqref{eqn:Radial_Wavenumber_q}), so it makes sense to drop the `$q$' label and just refer to radial basis functions by their radial wavenumber $k_R$:
\begin{equation}
    u^q(R) = u(k_R)\me^{ik_R R},
    \label{eqn:WKB_basis}
\end{equation}
where $u(k_R)$ is generally complex, and we suppress the slow $R_\mathrm{g}$ dependence. Note that we have \textit{not} assumed the fluctuations are tightly wound. We treat $k_R$ as a continuous variable, meaning we replace 
\begin{equation}
    \sum_q \to \frac{L}{2\pi}\int^{L/2}_{-L/2} \md k_R \equiv \fint \md k_R
    \label{eqn:sum_to_integral}
\end{equation}
for some reference length scale $L$ which is much larger than $\vert k_R \vert^{-1}$. Also, the coefficients $B^p(t) = B^{\ell}(k_R,t)$ are now distributions over $k_R$, i.e., we replace \eqref{eqn:phi_basis} with
\begin{equation}
    \delta \phi(\br, t) = \sum_{\ell} \fint \md k_R \,B^\ell(k_R, t)\,u(k_R) \,\me^{i(\ell\varphi+ k_RR)},
    \label{eqn:potential_fluctuation_WKB_expansion}
\end{equation}
and similarly for $\delta \phi^\mathrm{ext}$. In Appendix \ref{sec:WKB_basis_elements}, we show how to construct appropriate surface density basis functions in this regime and hence how to expand an arbitrary $\delta \Sigma(\br, t)$ in analogous form to \eqref{eqn:potential_fluctuation_WKB_expansion}.

The Volterra equation \eqref{eqn:Volterra_Equation} now becomes 
\begin{align}
     & B^\ell(k_R, t) = B_\mathrm{kin}^\ell(k_R, t) \nn 
     \\
     &\,\,\,\,+ \fint \md k_R' \int_{0}^t \md t' \mathcal{M}^\ell(k_R, k_R', t-t') \nn
     \\
     &\,\,\,\,\,\,\,\,\,\,\,\,\,\,\,\,\,\,\,\,\,\,\,\,\,\,\,\,\,\,\,\,\,\,\,\,\,\,\,\,\,\,\times [B^\ell(k_R', t')+B_\mathrm{ext}^\ell(k_R', t')],
     \label{eqn:Volterra_Equation_Short}
\end{align}
where 
\begin{align}
    B_\mathrm{kin}^\ell(k_R, t) &= -\frac{(2\pi)^2}{\mathcal{E}}\sum_{n_R}\int \md \bJ\, u_{\ell n_R}^q(\bJ)\nn
    \\
    &\,\,\,\,\,\,\,\,\,\,\,\,\,\, \times  
    \delta f_{\ell n_R}(\bJ, 0)\,\me^{-i(\ell\Omega_\varphi + n_R\Omega_R)\tau}.
    \label{eqn:Bkin_WKB}
 \end{align}
In Appendix \ref{sec:WKB_Kernel} we show that for a Schwarzschild DF \eqref{eqn:Schwarzchild}, the Volterra kernel in this regime reads
\begin{align}
    \mathcal{M}^{\ell}(k_R, k_R', \tau) = 
    \frac{1}{\mathcal{E}} 
    \int & \md J_\varphi [u(k_R)]^*u(k_R') 
    \nn
    \\
    & \times 
    G^{\ell}(k_R,k_R',\tau)\me^{-i\ell \Omega\tau},
    \label{eqn:Volterra_Kernel_Short}
\end{align} 
where the function $G^{\ell}(k_R,k_R',\tau)$ is defined in \eqref{eqn:aux_G}.

\begin{figure}
    \centering
        \includegraphics[width=0.99\linewidth]{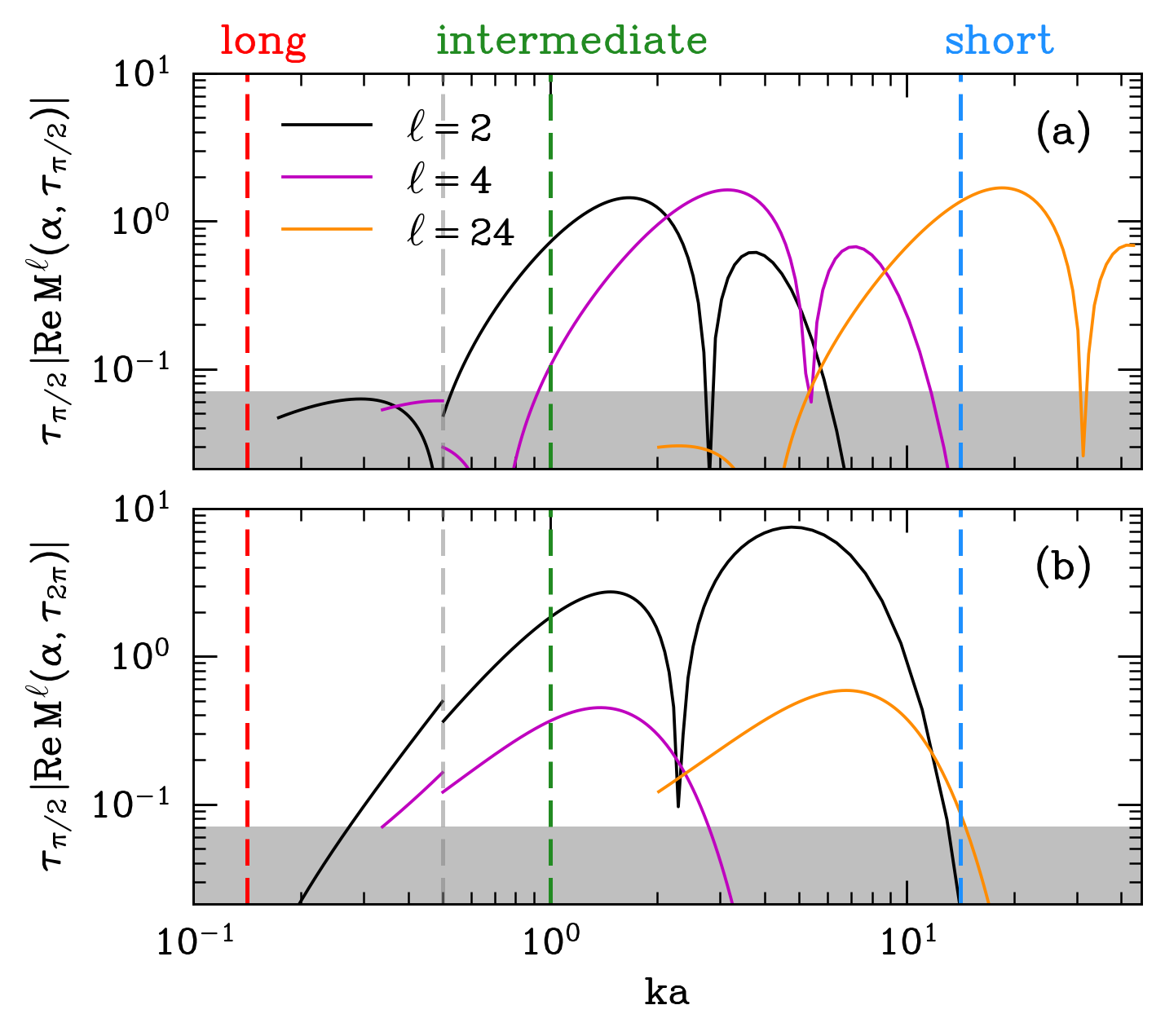}
    \caption{Magnitude of the Volterra kernel for the same log-spiral basis we discussed in \S\ref{sec:Linear_Theory_Example}. Panel (a) shows the value of $\tau_{\pi/2} \vert \mathrm{Re}\,\mathcal{M}^\ell(\alpha, \tau_{\pi/2})\vert$ for fixed $\ell$ and an active fraction $Y=0.1$ (see equation \eqref{eqn:active_fraction}),
    as a function of $ka$ (achieved by varying the pitch angle $\alpha$).
    The vertical dashed colored lines show the formal definitions of the asymptotic wavelength regimes; the light gray vertical dashed line shows the value $ka=0.5$ at which we transition from the long wavelength approximation \eqref{eqn:Volterra_Kernel_Long} to the short wavelength approximation \eqref{eqn:Volterra_Kernel_Short}.
    The gray band corresponds to values $\leq \epsilon$ (equation \eqref{eqn:small_kernel}). Panel (b) shows the same thing except the argument of the kernel is now $\tau_{2\pi}$.}
    \label{fig:Volterra_Kernel_Illustration_Wavelength_Dependence}
\end{figure}

We argued in \S\ref{sec:Characteristic_Scales} that self-gravity should be suppressed on very small scales, because it will be overwhelmed by velocity dispersion. We can now justify this mathematically by inspecting the short-wavelength Volterra kernel \eqref{eqn:Volterra_Kernel_Short}, and in particular the function $G^\ell$ (equation \eqref{eqn:aux_G}) that sits inside its integrand. The limit of very short wavelengths means $K_Ra, K_R'a \gg 1$ in \eqref{eqn:aux_G}; the factor $\me^{-(\mathcal{K}a)^2/(4\zeta)}$, with $\mathcal{K}$ given in \eqref{eqn:calK}, then guarantees that $G^\ell$ (and hence $\mathcal{M}^\ell$) is strongly suppressed at most $\tau$. The exceptions are during the short time windows when $\mathcal{K}a\lesssim 1$, but from \eqref{eqn:calK}-\eqref{eqn:time_dependent_angle} we see that these windows last $\ll \kappa^{-1}$.

As an example, in panel (a) of Figure \ref{fig:Volterra_Kernel_Illustration_Wavelength_Dependence} we plot $\tau_{\pi/2} \, \vert \mathrm{Re}\,\mathcal{M}^\ell(\alpha, \tau_{\pi/2})\vert$ for the same log-spiral basis used in \S\ref{sec:Linear_Theory_Example}, as a function of $ka$. To vary $k$, we fix the value of $\ell$ and vary $\alpha \in (0.1^\circ, 89.9^\circ)$ (see equation \eqref{eqn:nearlyslogspiral_wavenumber_modulus}). Again, we choose an active fraction of $Y=0.1$ (equation \eqref{eqn:active_fraction}). The different solid colored lines correspond to different $\ell$ values as indicated in the legend; the red, green and blue vertical dashed lines correspond to the formal definitions of the long, intermediate and short wavelength regimes respectively, while the gray shaded region has height $\epsilon$, i.e. it corresponds to negligible self-gravity according to equation \eqref{eqn:small_kernel}. In creating this figure, we have glued together results from the long and short wavelength regimes: the light gray vertical dashed line indicates $ka = 0.5$, below which we use equation \eqref{eqn:Volterra_Kernel_Long} and above which we use equation \eqref{eqn:Volterra_Kernel_Short}. Panel (b) of the Figure shows the same quantities as panel (a), but with the argument of the Volterra kernel replaced with $\tau_{2\pi}$.

We see from these panels that for small $\ell$, the kernel is maximized in the vicinity of intermediate wavelengths, $ka\sim $ a few, although the precise scale at which it is maximized depends strongly on $\ell$ and $\tau$. On the other hand, for larger $\ell$, the early-time ($\sim \tau_{\pi/2}$) kernel tends to peaks at larger values of $ka$ --- see the orange line in Figure \ref{fig:Volterra_Kernel_Illustration_Wavelength_Dependence}a. However, this does not imply a strong self-gravitating response at short wavelengths, because the large value of $\vert\mathcal{M}^\ell\vert$ is very short-lived: it drops precipitously as $\tau$ increases, as we see from Figure \ref{fig:Volterra_Kernel_Illustration_Wavelength_Dependence}b. This is consistent with the behavior we saw in the rightmost column of Figure \ref{fig:Volterra_Kernel_Illustration}, where decreasing $\alpha$ (increasing $ka$) gave rise to a larger initial `pulse' in $\mathcal{M}^\ell$, but then strongly suppressed any further oscillations. Indeed, in all the examples shown here self-gravity is negligible at the shortest wavelengths beyond $\tau_{2\pi} = T_8/\vert \ell \vert$, which is often a very small fraction of an orbital time.

\begin{figure}
    \centering
    \includegraphics[width=0.99\linewidth]{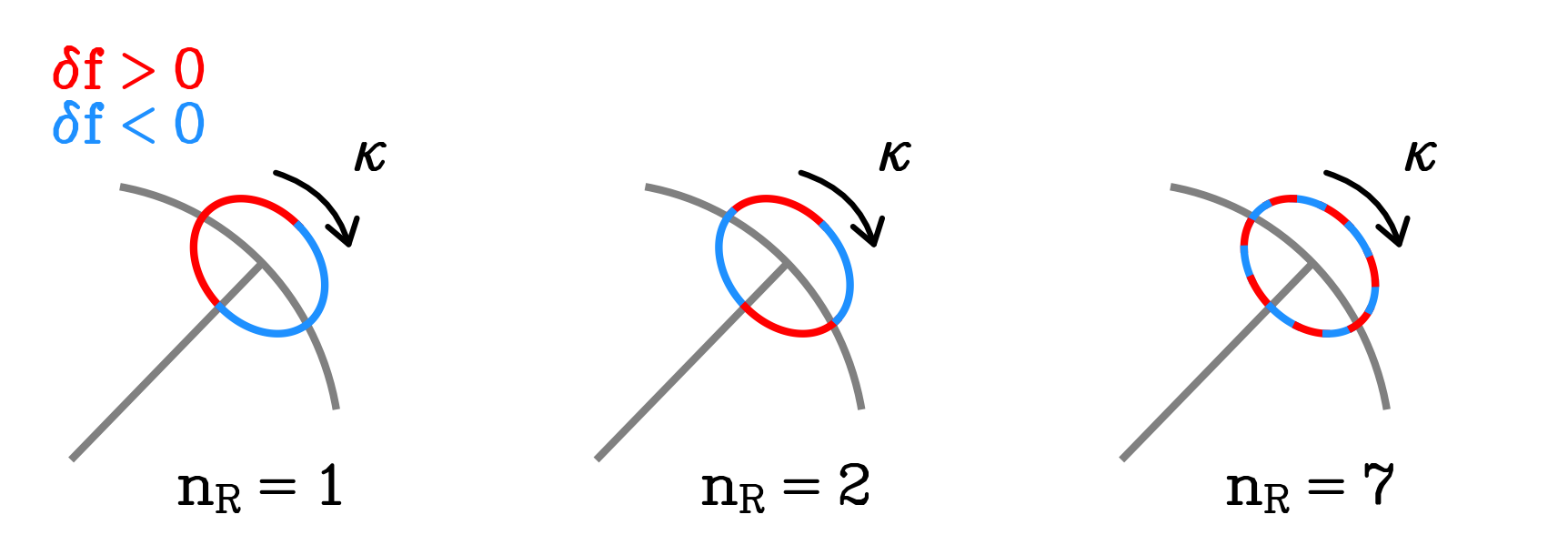}
    \caption{Schematic diagram of of different $n_R$ components of a DF fluctuation $\delta f$ at a fixed value of the actions $\bJ$. These all rotate around the epicycle at the same `pattern speed' $\kappa$ but have widely differing effective frequencies $n_R\kappa$.}
    \label{fig:sketch_nR}
\end{figure}

Physically, though, it is not immediately obvious why this should be the case. We tend to think of $\sigma$ as `random velocity', but in fact it simply encodes the amplitude of radial epicyclic motions, which all occur roughly in phase (i.e., with a single frequency $\kappa$) for stars with comparable $J_\varphi$. The reason velocity dispersion acts to kill self-gravity is that on small scales, the fluctuations have power in Fourier components with many different $n_R$ values --- see Figure \ref{fig:sketch_nR} --- and all these Fourier components `rotate' around their epicyclic ellipses at different rates $n_R\kappa$, even at fixed $\bJ$. The beating of these many different oscillations means that DF fluctuations cannot talk to one another coherently over the timescale required for their self-gravity to become important, which we argued in \S\ref{sec:Characteristic_Scales} was $\gtrsim\Omega^{-1}$.
Indeed, it is sometimes said that the effect of finite velocity dispersion can be mimicked by considering a cold disk but including gravitational softening, e.g., replacing Newton's interaction kernel \eqref{eqn:Newtonian_Kernel} with a Plummer kernel with a scale radius $r_\mathrm{s} \sim a$.
In reality, the role of velocity dispersion is to take the kernel $\mathcal{M}^{pp'} \to 0$ on small scales and hence reduce the power spectrum of fluctuations below the length scale $a$ to the level of Poisson noise (rather than the exponential suppression that would be implied by softening). Of course, finite-$N$ noise is very low in real galaxies, and even swing-amplified Poisson noise would still fall far below the amplitude requirements from \S\ref{sec:amplitude_requirements}, so this level is effectively zero.


\subsection{Large-$m$ limit}
\label{sec:Small_Azimuthal}

Let us consider spirals with a large number of arms --- formally we assume $m\sim \epsilon^{-2}$, so that we are guaranteed to be in the short-wavelength regime regardless of the value of $k_R$ (Paper I).\footnote{Though the theory we are about to develop is often applied far beyond this formal regime of validity (e.g., \citealt{toomre1981amplifies}).} Then the Volterra equation is given in equation \eqref{eqn:Volterra_Equation_Short} with the kernel \eqref{eqn:Volterra_Kernel_Short}.

We now proceed to recover the classic shearing sheet Volterra equation of \citep{julian1966non} (see also \citealt{Fuchs2001-ko,binney2020shearing}), starting with \eqref{eqn:Volterra_Kernel_Short} and listing our assumptions as we go. Any of these assumptions could be relaxed, and the resulting Volterra equation would constitute a generalization of the shearing sheet results of \cite{julian1966non}, though we will not pursue any such generalization here.

We focus on an annulus centered at some radius $R_*$, corresponding to an angular momentum for a circular orbit of $J_\varphi(R_\mathrm{g} =R_*) \equiv J_*$. We express an arbitrary angular momentum and guiding radius of a star within this annulus as in equation \eqref{eqn:expanding_R_and_J}. Then in considering \eqref{eqn:Volterra_Kernel_Short} (and in particular the function $G^\ell$ in equation \eqref{eqn:aux_G}) we will treat $R_*, J_*$ as `slow' variables and $r,j$ as `fast' variables. The wavenumbers $K_R$ and $K_R'$ only vary on the slow scale $\sim J_*$ (see equation \eqref{eqn:modified_k}). We also assume that our DF and rotation curve are rather featureless, so that $g_0$ and $\langle J_R \rangle$ also vary on the slow scale $\sim J_*$; then terms involving their derivatives are negligible in \eqref{eqn:aux_G}.

Proceeding with these assumptions, we get a much simpler version of the function $G^\ell$ from equation \eqref{eqn:aux_G}, which we plug into \eqref{eqn:Volterra_Kernel_Short}, with the result
\begin{align}
    &\mathcal{M}^\ell(k_R,k_R',\tau) \simeq 
    \frac{2\pi^2}{\mathcal{E}} 
    [u(k_R)]^*\,u(k_R') \,g_0 \, a^2 {K_RK_R'}
    \nn
    \\
    &\,\,\,\,\,\,\,\,\times
    \int_{-\infty}^\infty \md j\, 
    \me^{-i(k_R-k_R')\Rg - i\ell\Omega\tau}
    \frac{\sin \theta}{\zeta^2}\,
    \me^{-(\mathcal{K}a)^2/(4\zeta)}.
    \label{eqn:Volterra_Kernel_Sheet_Intermediate}
\end{align}
The integrand in this expression is a function of five $j$-dependent, dimensionless quantities. Three of these are explicit, namely $k_RR_\mathrm{g}$,  $k_R'\Rg$ and $\Omega\tau$. In addition, the factor $\zeta^{-2}{\sin \theta}\,\me^{-(\mathcal{K}a)^2/(4\zeta)}$ also depends on $\langle \Omega_\mathrm{D}\rangle\tau$ (see equation \eqref{eqn:Dehnen_phase}) and $\kappa\tau$. The next assumption we need is that the dominant $j$ dependence in this integrand can be captured by simply expanding the explicit phase
\begin{align}
    (k_R-k_R') R_\mathrm{g} + &\ell\Omega\tau  \simeq (k_R-k_R') R_* + \ell\Omega_*\tau   \nn
    \\
    &\,\,\,\, + (k_R-k_R')\frac{\md R_*}{\md J_*}j + \ell \frac{\md \Omega_*}{\md J_*}j\tau,
    \label{eqn:expanding_the_phase}
    \end{align}
and ignoring all other $j$ dependence.
With this we can carry out the $j$ integral in \eqref{eqn:Volterra_Kernel_Sheet_Intermediate} using $\int_{-\infty}^\infty \md j\, \me^{isj} = 2\pi \delta(s)$. We find
\begin{align}
    &\mathcal{M}^\ell(k_R,k_R',\tau) \simeq 
    -\frac{8\pi^3}{\mathcal{E}} 
    [u(k_R)]^*\,u(k_R') \,g_0 \, a^2 R B {K_RK_R'} 
    \nn
    \\
    & \times \frac{\sin \theta}{\zeta^2} \, \me^{-(\mathcal{K}a)^2/(4\zeta) -i(k_R-k_R')R - i\ell\Omega \tau} 
    \nn
    \\
    & \times
    \delta[k_R'-(k_R -2A k_\varphi^\ell\tau)],
    \label{eqn:Volterra_Kernel_Sheet_Dehnen}
\end{align}
where $  B \equiv A-\Omega$ is the Oort $B$ constant (see equations \eqref{eqn:Oort_Constants}), and we have used the definition \eqref{eqn:Azimuthal_Wavenumber_q}, and everything is to be evaluated at $R=R_*$.

Physically, equation \eqref{eqn:Volterra_Kernel_Sheet_Dehnen} (coupled with the Volterra equation \eqref{eqn:Volterra_Equation_Short}) tells us that the only wavenumbers that affect the fluctuation with wavenumber $k_R$ today (time $t$) are those wavenumbers $k_R'$ from times $t-\tau$ in the past that would evolve to the value $k_R$ today by simple shear, namely $k_R' = k_R -2A k_\varphi^\ell \tau$. This result is trivial if one \textit{starts} from the equations of the sheared sheet, since in this case the problem is homogeneous when viewed in sheared coordinates \citep{goldreich1965ii,julian1966non}, and in the corresponding linear theory there is no cross-talk between different (time-dependent) Fourier wavenumbers.
Here, however, we have \textit{derived} this result by starting from a much more general global linear theory.

\subsubsection{Dehnen drift and the shearing sheet}

In fact, equation \eqref{eqn:Volterra_Kernel_Sheet_Dehnen} is still more general than the usual shearing sheet theory because it accounts for the Dehnen drift, encoded in the function $\zeta$ (equation \eqref{eqn:Dehnen_phase}). Ignoring the Dehnen drift means setting $\zeta = 1$,  which, if we want errors to our theory to be $\mathcal{O}(\epsilon^2)$ or smaller, formally requires
\begin{equation}
    \bigg \vert \ell \langle J_R \rangle \frac{\md \kappa}{\md J_\varphi} \tau \bigg \vert \lesssim \epsilon^2,
    \label{eqn:ignore_Dehnen_sheet}
\end{equation}
for all relevant $\tau$ in \eqref{eqn:Volterra_Equation_Short}.

In a typical galaxy $\vert \md \kappa /\md J_\varphi \vert \sim \kappa/J_\varphi$, meaning the above inequality becomes
\begin{equation}
     m  \kappa  \tau  \lesssim 1.
    \label{eqn:ignore_Dehnen_sheet_1}
\end{equation}
Since we have formally ordered $m =\vert \ell \vert \sim \epsilon^{-2}$ in this section, the inequality \eqref{eqn:ignore_Dehnen_sheet} is correct only for a very short time,  $\tau \lesssim \epsilon^2 \kappa^{-1}$. Of course, it is possible to relax these orderings somewhat if we are willing to allow errors $\mathcal{O}(\epsilon)$ into the theory. But whatever formal error threshold we set, the Dehnen drift cannot be ignored once $m\tau$ is large enough.

The shearing sheet approximation avoids this issue entirely by insisting on $\md \kappa/\md J_\varphi = 0$ from the start. In that case, the Dehnen drift \eqref{eqn:Dehnen} is identically zero and the inequality \eqref{eqn:ignore_Dehnen_sheet} is satisfied at all $\tau$. The shearing sheet Volterra kernel therefore reads
\begin{align}
    &\mathcal{M}^\ell(k_R,k_R',\tau) \simeq 
    -\frac{8\pi^3}{\mathcal{E}} 
    [u(k_R)]^*\,u(k_R') \,g_0 \, a^2 R B {K_RK_R'}
    \nn
    \\
    & \times {\sin \theta} \, \me^{-(\mathcal{K}a)^2/(4) -i(k_R-k_R')R - i\ell\Omega \tau} 
    \nn
    \\
    & \times
    \delta[k_R'-(k_R -2A k_\varphi^\ell\tau)],
    \label{eqn:Volterra_Kernel_Sheet}
\end{align}
However, this cannot be rigorously justified. In other words, there is no consistent scheme of approximations (errors $\mathcal{O}(\epsilon^2)$) to the general global equations of motion that will reproduce the equations of motion normally assumed in the shearing sheet. Exploring the consequences of this fact for phenomena like swing amplification (\S\ref{sec:swing_amplification}) is a clear avenue for future work, but is beyond the scope of this paper. 

\subsubsection{Swing amplification}
\label{sec:swing_amplification}

In Appendix \ref{sec:Recover_Binney} we show explicitly that the Volterra equation \eqref{eqn:Volterra_Equation_Short} with the shearing sheet kernel \eqref{eqn:Volterra_Kernel_Sheet} truly does match what \cite{binney2020shearing} calls the `JT equation' after \cite{julian1966non}. This confirms that our theory will recover the results of the classic shearing sheet that have been explored extensively by \cite{julian1966non,Fuchs2001-ko,binney2020shearing} and others.

While we will not rederive (or re-explain) their results here, the most important outcome of the JT equation  is \textit{swing amplification}, which famously allows for small-amplitude fluctuations in the disk potential and surface density to be boosted, at least temporarily (on an orbital timescale $\sim \Omega^{-1}$). The maximum amplification factor of an $m$-armed spiral (azimuthal wavelength $\lambda_\varphi = 2\pi R_*/m$) under swing amplification (e.g., Figure 6 of \citealt{binney2020shearing}) is reasonably well-fit, in the region where the amplification is physically important (see \S\ref{sec:amplitude_requirements}), by a Gaussian
\begin{align}
    \mathcal{A}_\mathrm{max}\simeq W(Q) \times \exp \left( \frac{[\lambda_\varphi - 1.25\lambda_\mathrm{crit}(Q)]^2}{2 \times [0.5\lambda_\mathrm{crit}(Q)]^2}\right),
    \label{eqn:amplification_max}
\end{align}
where $\lambda_\mathrm{crit} = 2\pi/k_\mathrm{crit}$ and
\begin{equation}
    k_\mathrm{crit} \equiv \frac{\kappa^2}{2\pi G\Sigma} \simeq \frac{0.756\,Q}{a},
    \label{eqn:critical_wavenumber}
\end{equation} 
is the so-called \textit{critical wavenumber}. The function $W$ is a sharply declining function of $Q$; for instance, $W(1.2)\simeq 100$ whereas $W(2)\simeq 4$; larger values of $Q$ are irrelevant (see \S\ref{sec:amplitude_requirements}).

This is consistent with the argument of \S\ref{sec:Characteristic_Scales} that self-gravity should matter most when $Q\sim 1$ and $ka \sim 1$.
However, the remarkable thing about swing amplification is that many e-foldings of amplification are achieved on a timescale $\sim \Omega^{-1}$, whereas for self-gravitating collapse (Jeans instability) we expect about one $\me$-folding per orbital time (e.g., \S\ref{sec:max_growth_rate}). The reason this is possible is that swing amplification is not a true instability, but rather a temporary coming together of phases of epicyclic orbits. While the real-space imprint of this `phase unmixing', namely the corresponding density/potential fluctuation, can be very large at peak amplitude, any such episode will necessarily be short-lived (though see \S\ref{sec:Nonlinear}).

\subsection{Tightly wound limit}
\label{sec:Small_Radial}

Let us now consider short-wavelength fluctuations in the tightly wound limit. Formally, we defined this in Paper I to be $\vert k_\varphi/k_R\vert \sim\epsilon^2$, which required a small number of arms $m\sim 1$ and large radial wavenumber $\vert k_R\vert  a\sim\epsilon^{-1}$. Thus we are in a regime distinct from \S\ref{sec:Small_Azimuthal}, which formally relied on a large $m \sim \epsilon^{-2}$ but placed no restriction on $\vert k_R\vert$. Yet once again we will find that the connection between the two regimes is smooth.

In the tightly wound limit, we can simplify the short-wavelength kernel \eqref{eqn:Volterra_Kernel_Short} by setting  $K_R \to \vert k_R\vert $, $K_R' \to \vert k_R'\vert $, and $\beta, \beta' \to 0$. We can then proceed as in \S\ref{sec:Small_Azimuthal}, making the same assumptions up to the expansion \eqref{eqn:expanding_the_phase} with these replacements. However, we now also assume that the term proportional to $(k_R-k_R')$ in \eqref{eqn:expanding_the_phase} dominates.\footnote{This is fine since the ratio of the two terms is at most $\sim \vert\ell\Omega \tau_\mathrm{conv} \vert /\vert k_R R_\mathrm{g} \vert$, where $\tau_\mathrm{conv}$ is the convergence time we introduced in equation \eqref{eqn:small_kernel}. We know from \S\ref{sec:Linear_Theory_Example} that typically $\tau_\mathrm{conv} \lesssim \Omega^{-1}$ so that the ratio is $\sim \vert \ell / k_RR_\mathrm{g}\vert = \vert k_R/k_\varphi\vert \sim \epsilon^2$ by the assumption of tight winding.} Then in particular, the delta function in \eqref{eqn:Volterra_Kernel_Sheet} is replaced by $\delta(k_R'-k_R)$. Putting these pieces together and using the identity $\me^{z\cos\theta}= \sum_{n=-\infty}^\infty I_n(z)\me^{in\theta}$, instead of \eqref{eqn:Volterra_Kernel_Sheet} we get 
\begin{align}
    &\mathcal{M}^\ell(k_R,k_R',\tau)  \simeq 
    \frac{4\pi^3i}{\mathcal{E}} \vert u(k_R) \vert^2 g_0 RB (k_Ra)^2 \delta(k'_R-k_R)
    \nn
    \\
    &\times
    \me^{-(k_Ra)^2/2}
    \sum_{n=-\infty}^\infty I_n(\chi)  \left[ \me^{i(n+1)\kappa\tau} - \me^{i(n-1)\kappa\tau}  \right],
    \label{eqn:Volterra_Kernel_Short_Radial_Wavelength}
\end{align}
where
\begin{equation}
    \chi \equiv \left( \frac{k_R\sigma}{\kappa}\right)^2 = \frac{(k_Ra)^2}{2}.
\end{equation}
Of course, this is precisely the kernel we \textit{would} have found had we taken the shearing sheet kernel \eqref{eqn:Volterra_Kernel_Sheet}, derived for large $m$, and then extrapolated the result to small $m$.

The terms in \eqref{eqn:Volterra_Kernel_Short_Radial_Wavelength}  are manifestly  sourced by all possible `Lindblad-resonance' frequencies. However, there is no surviving `corotation' term. This is a famous consequence of taking the tightly wound limit, which excludes most of the interesting physics (including swing amplification) that was possible in \S\ref{sec:Small_Azimuthal}.

\subsubsection{Lin-Shu-Kalnajs modes}
\label{sec:LSK}

The major advantage of the reduced kernel \eqref{eqn:Volterra_Kernel_Short_Radial_Wavelength}
is that it makes the integral over $k_R'$ in equation \eqref{eqn:Volterra_Equation_Short} trivial: there is no longer any cross-talk between distinct basis functions. This simplification makes it possible to perform many explicit calculations that are not tractable otherwise. The most important such calculation is the dispersion relation for Lin-Shu-Kalnajs (LSK) modes. To derive these we first Laplace transform \eqref{eqn:Volterra_Kernel_Short_Radial_Wavelength} (see equation \eqref{eqn:Laplace}), then use the Landau dispersion relation \eqref{eqn:Landau_Dispersion_Relation}, substitute the relations \eqref{eqn:curlyE_WKB} (with $k=\vert k_R\vert$) and \eqref{eqn:Surface_Density_g0}, and use the identity $I_{n-1}(\chi)-I_{n+1}(\chi) = 2n I_n(\chi)/\chi$. The result is
\begin{align}
    1 =
    \frac{\vert k_R\vert}{k_\mathrm{crit}} \me^{-\chi} 
    \sum_{n=-\infty}^\infty \frac{n^2}{n^2 - (\omega/\kappa)^2} \frac{I_{n}(\chi)}{\chi},
    \label{eqn:LSK_Dispersion_Relation}
\end{align}
where $k_\mathrm{crit}$ was defined in \eqref{eqn:critical_wavenumber}. This is equivalent to the dispersion relation in equation (75) of \cite{binney2020shearing}\footnote{Note Binney uses a different convention for Laplace transforms than us; his $p$ is equal to our $-i\omega$, meaning his $s$ is equal to our $\omega/\kappa$.}.

The key idea behind a lot of tightly wound spiral structure theory, including the Lin-Shu theory of `quasi-stationary spiral structure', is that that the general solution to \eqref{eqn:Volterra_Equation_Short} can be written as a superposition of these LSK modes --- essentially, these are the van Kampen modes (\S\ref{sec:van_Kampen}) of the tightly wound limit of the linear response problem. There are, however, several issues with this approach, as we discuss briefly in \S\ref{sec:previous_literature}.
\section{Discussion}
\label{sec:Discussion}

In this paper, we have considered the linear theory of spiral structure in stellar disks (\S\ref{sec:Linear}), which we specialized to various asymptotic wavelength regimes (\S\S\ref{sec:Long_Wavelengths}-\ref{sec:Short_Wavelengths}). In doing so we recovered (and in some cases generalized) several classic results, including Lindblad-Kalnajs density waves (\S\ref{sec:Long_Wavelengths_Weak_Self_Gravity}), groove instabilities (\S\ref{sec:Long_Wavelength_Instability_Sharp}), swing amplification (\S\ref{sec:swing_amplification}), and LSK modes (\S\ref{sec:LSK}). We also found that these results tend to blend into one another, i.e., our asymptotic formulae often connect smoothly when extrapolated beyond their natural regimes of validity.

The fact that asymptotic results connect smoothly is, on the one hand, a great benefit to spiral structure theorists, since it renders analytic theory useful over a wider range of scales than might have been expected. On the other hand, it can obscure the \textit{distinctions} between different spiral theories that are superficially similar. Meanwhile the representation degeneracy mentioned in \S\ref{sec:choice_of_representation} and Appendix \ref{sec:App_Representations} has tended to obscure the \textit{similarities} between spiral theories that are superficially distinct.

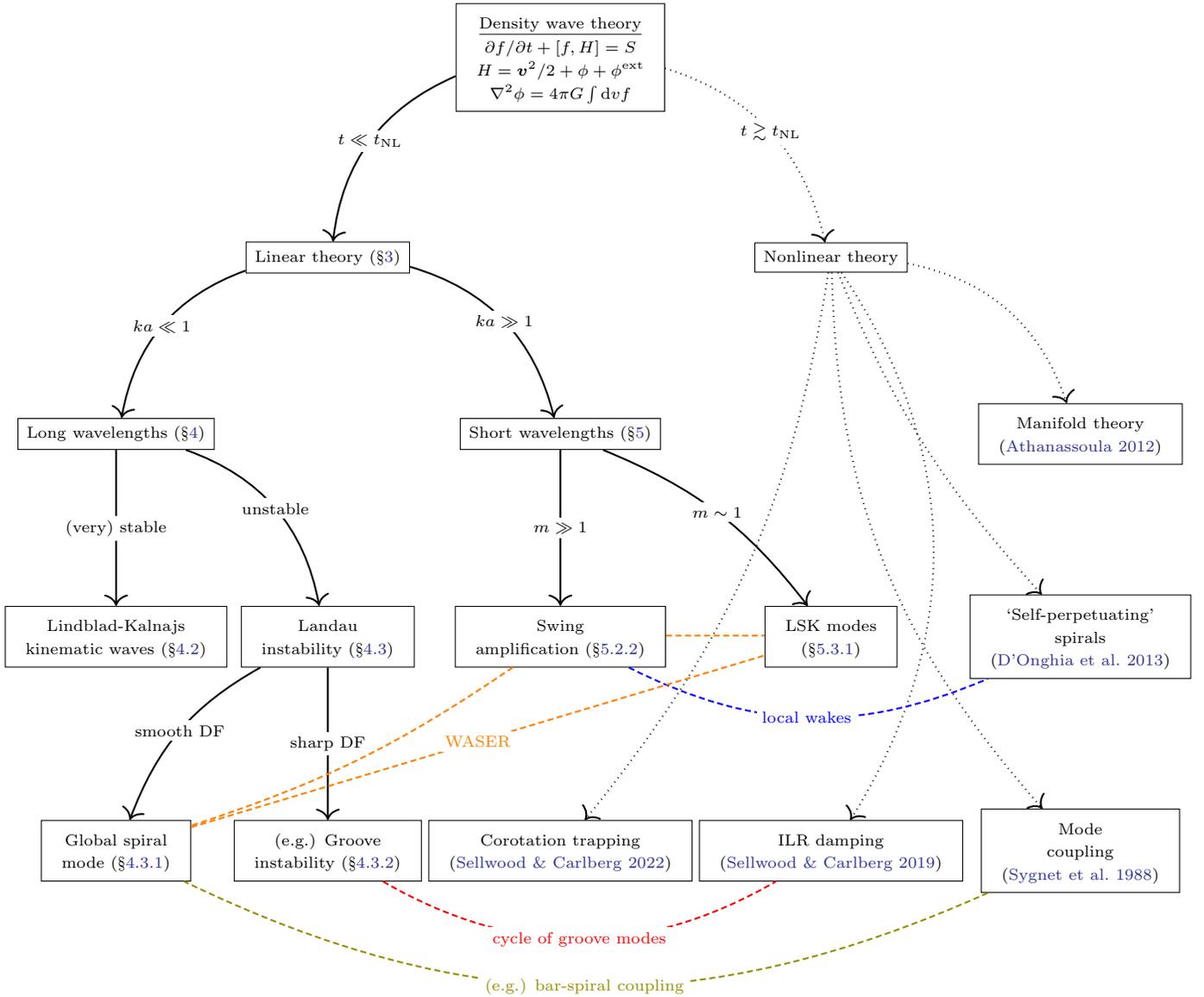
\begin{figure*}
    \begin{tikzcd}[cells={nodes={draw, opacity=1}, box/.style={draw,fill=white,align=center}, font = {\scriptsize}}, row sep=2cm, column sep=0.1cm]
    & &  \begin{tabular}{c}
    \underline{Density wave theory}
    \\
    $\p f/\p t + [f,H] = S$ \\
    $H=\bm{v}^2/2 + \phi+ \phi^\mathrm{ext}$ \\
    $\nabla^2 \phi = 4\pi G \int\md v f$
    \end{tabular}
    \arrow[thick, dl, bend right, smalltext= $t\ll t_\mathrm{NL}$]  \arrow[dotted, thick, dr, bend left, smalltext= $t\gtrsim t_\mathrm{NL}$]  \\
    &
\textup{Linear theory (\S\ref{sec:Linear})} 
\arrow[thick, dl, bend right , smalltext= $ka \ll 1$] \arrow[thick, dr, bend left, smalltext= $ka \gg 1$]
&
    & \textup{Nonlinear theory} \arrow[dotted, thick, dddl, bend left=15] \arrow[dotted, thick, ddd, bend left=33] \arrow[dotted, thick, dddr, bend right=20] \arrow[dotted, thick, ddr, bend right=10] \arrow[dotted, thick, dr, bend left]
%
%
\\
\textup{Long wavelengths (\S\ref{sec:Long_Wavelengths})}  
\arrow[thick, d, smalltext= (very) stable] 
\arrow[thick, dr, bend left, smalltext= unstable]
&
&
\textup{Short wavelengths (\S\ref{sec:Short_Wavelengths})} 
\arrow[thick,d, smalltext= $m\gg 1$] \arrow[thick,dr, bend left=15, smalltext= $m\sim 1$]
&
&
  \begin{tabular}{c}
Manifold theory \\
(\citealt{athanassoula2012manifold})
\end{tabular} 
\\
  \begin{tabular}{c}
Lindblad-Kalnajs \\
kinematic waves (\S\ref{sec:Long_Wavelengths_Weak_Self_Gravity})
\end{tabular} 
& 
  \begin{tabular}{c}
Landau \\
instability (\S\ref{sec:Global_Instabilities}) 
\end{tabular} 
\arrow[thick,dl, bend right=20 , smalltext= smooth DF]
\arrow[thick,d, smalltext= sharp DF]
    &   
      \begin{tabular}{c}
Swing \\
amplification
(\S\ref{sec:swing_amplification})
\end{tabular} 
\arrow[orange, dashed, thick, -, dll, bend left = 8
]
\arrow[orange, dashed, thick, -, r, bend left = 0
]
\arrow[blue, dashed, thick, -, rr, smalltext= local wakes, bend right = 25
]
& 
  \begin{tabular}{c}
LSK modes \\
(\S\ref{sec:LSK})
\end{tabular} 
&   \begin{tabular}{c}
`Self-perpetuating' \\ spirals \\
\citep{d2013self}
\end{tabular} 
\\
  \begin{tabular}{c}
Global spiral \\
mode (\S\ref{sec:smooth_DF})
\end{tabular} 
\arrow[orange, dashed, thick, -, urrr, bend right = 0, smalltext=WASER
]
&
  \begin{tabular}{c}
(e.g.) Groove \\
instability (\S\ref{sec:Long_Wavelength_Instability_Sharp})
\end{tabular}  
\arrow[red, dashed, thick, -, rr, bend right = 30, smalltext=cycle of groove modes
]
  &  
  \begin{tabular}{c}
Corotation trapping \\
(\citealt{sellwood2022spiral})
\end{tabular} 
    & \begin{tabular}{c}
ILR damping \\ 
\citep{sellwood2019spiral}
\end{tabular} 
 &  
   \begin{tabular}{c}
Mode \\ coupling \\
(\citealt{Sygnet1988-eq})
\end{tabular} 
\arrow[olive, dashed, thick, -, llll, bend left = 25, smalltext = {(e.g.) bar-spiral coupling}
]
\end{tikzcd}
    \caption{Summary of density wave theories of spiral structure. Thick black lines point to aspects of (linear) density wave theory that we have covered in this paper, while dotted black lines denote (nonlinear) aspects that we have not considered. Colored dashed lines illustrate important links between various phenomena.}
    \label{fig:Summary}
\end{figure*}

The purpose of this section is to discuss how our work clarifies these issues, offering a framework for a true synthesis of the general theory of spiral structure in galaxies. In Figure \ref{fig:Summary} we provide a schematic illustration of what such a synthesis would look like. With black solid arrows we sketch  the logical chain of approximations that takes us from the general linear theory to specific spiral phenomena. We emphasize that in this work we have ignored nonlinear theory almost entirely:  a (non-exhaustive) collection of nonlinear ideas is indicated with dotted arrows. As we mentioned in the Introduction, many nonlinear theories are deeply connected with linear results, and we have attempted to illustrate some of these connections with colored dashed lines in Figure \ref{fig:Summary}. We have also ignored the role of gas in our theory, other than as an external driver of fluctuations. Many of the general ideas from linear theory also work in the gaseous case \citep{Binney2008-ou}, but there are certainly important gas-only phenomena (both linear and non-linear), as well as non-trivial couplings between stars and gas \citep{kim2007gravitational}, that would alter the picture developed here.

Importantly, most observed spiral structure is of intermediate wavelength, $ka\sim 1$ (see \S\ref{sec:geometry}). Mathematically, we can investigate the intermediate regime by extrapolating the kernels from the long or short wavelength regimes, i.e., by taking the $ka\gg 1$ limit of \eqref{eqn:Volterra_Kernel_Long} or by taking the $k\Rg \ll 1$ limit of \eqref{eqn:Volterra_Kernel_Short}. In either case we arrive at the same answer, namely an expression the same as \eqref{eqn:Volterra_Kernel_Long} but with the replacement\footnote{In Figure \ref{fig:Volterra_Kernel_Illustration}, the smooth transition through the value \eqref{eqn:Volterra_Equation_Intermediate} occurred somewhere in the range $ka\in(0.3, 0.5)$, but in general the precise transition value depends on the precise basis functions and parameters chosen.} 
\begin{equation}
    \xi_{\pm}^{\ell qq'}(\Rg, a) \to  - \frac{ a^2}{4}  [k^q_R]^* k^{q'}_R,
    \label{eqn:Volterra_Equation_Intermediate}
\end{equation}
in the definition \eqref{eqn:aux_pm} of the auxiliary functions $G^{\ell q q'}_{\pm}(J_\varphi, \tau)$, and we identify $u^q(R_\mathrm{g}) = u(k_R)$ and $u^{q'}(R_\mathrm{g}) = u(k_R')$. However, as we showed in Paper I, errors in the intermediate regime are formally $\mathcal{O}(\epsilon)$. As a result, we do not expect many of the asymptotic results that we derived in the short/long wavelength regimes to hold up to the desired $\mathcal{O}(\epsilon^2)$ accuracy when applied to real galaxies; one may have to settle for $\mathcal{O}(\epsilon)$ instead.

Moreover, since the asymptotic wavelength limits are largely what \textit{define} the different spiral theories (Figure \ref{fig:Summary}), in some cases it may not be profitable to try to distinguish `the' mechanism that caused a particular spiral (either in observation or simulation). For instance, whether some ragged, intermediate-wavelength spiral with $m=4$ arms\footnote{This is what we believe the Milky Way's spirals look like \citep{vallee2017guided,Eilers2020-na}.} is a result of swing amplification (\S\ref{sec:swing_amplification}) or LSK modes (\S\ref{sec:LSK}) might not be a well-posed question, since it exists in a regime where neither concept is formally well-defined, and  in such circumstances any theory that relies on these concepts may not be falsifiable. On the other hand one \textit{should} be able to distinguish, at least in simulations, temporary amplification of noise on the one hand and exponentially-growing global modes on the other. The groove instability cycle of Sellwood \& Carlberg (see \S\ref{sec:Nonlinear}) is appealing in this context since it is falsifiable (at least numerically, but perhaps even observationally --- see \citealt{Sellwood2019-xd}).

The rest of this section is dedicated to fleshing out in prose what is illustrated diagrammatically in Figure \ref{fig:Summary}. We begin in \S\ref{sec:previous_literature} by explaining where many classic papers on linear spiral structure theory fit into our scheme. Then, in \S\ref{sec:Nonlinear}, we turn to nonlinear theories and some of the open questions associated with them.


\subsection{Relation to previous literature}
\label{sec:previous_literature}

Although we will proceed in this section in a vaguely chronological order, it is certainly not intended as an exhaustive literature review, still less an accurate history \citep{pasha2004density,pasha2004density2,marochnik2005west}. It is especially incomplete as regards spirals in gaseous disks. We refer the reader to, e.g., Chapter 6 of \cite{Binney2008-ou} and the review by \cite{SellwoodMasters} where many more references can be found.

The quantitative theory of spiral structure really started with the axisymmetric stability analysis of \cite{toomre1964gravitational}, which follows from the $m\to 0$ limit of the tightly wound results of \S\ref{sec:Small_Radial}. There soon followed the local nonaxisymmetric stability and modal analysis of \cite{Lin1964,Kalnajs1965} whose central result is the LSK dispersion relation \eqref{eqn:LSK_Dispersion_Relation}.

The next major step was again based on local theory, this time in the shearing sheet. \cite{julian1966non}, using the Volterra representation of linear theory (\S\ref{sec:Linear}), showed that a mass placed on a circular orbit in a (weakly) stable stellar disk would produce a `swing amplified' spiral-shaped wake. Their results all follow from the kernel \eqref{eqn:Volterra_Kernel_Sheet} --- see \S\ref{sec:swing_amplification} and Appendix \ref{sec:Recover_Binney}.

Explanations, reformulations, and elaborations of the swing amplifier  have been the focus of many papers since \cite{julian1966non} (and its gaseous precursor, \citealt{goldreich1965ii}). In the gaseous context, \cite{Goldreich1978-xe} studied linear perturbations in the shearing sheet and showed that these satisfy an oscillator equation, with the `spring constant' of the oscillator being time dependent and swing amplification being attributed to the spring constant temporarily becoming negative. Soon after, \cite{toomre1981amplifies} studied the stellar shearing sheet and derived an analogous oscillator equation for the Lagrangian (see \S\ref{sec:Lagrangian}) displacement of stellar trajectories.\footnote{However, \cite{toomre1981amplifies} had to assume his stars had zero velocity dispersion in order to make the Lagrangian calculation analytically tractable. He then fudged the effect of finite velocity using the LSK reduction factor.} Toomre's Lagrangian interpretation  has since been studied further by \cite{Michikoshi2016-kq,michikoshi2016swing,Michikoshi2020-yh,yoshida2021elementary,yoshida2023phase}. An alternative derivation of the \cite{julian1966non} results, this time using the Landau representation (\S\ref{sec:Landau}), was given by \cite{Fuchs2001-ko}. Yet another derivation, this time back in the Volterra language, was performed by \cite{binney2020shearing}, who also extended and clarified some of the original 1966 theory. All the analytic results mentioned in this paragraph can be derived from the kernel \eqref{eqn:Volterra_Kernel_Sheet}. However, they are typically derived \textit{starting} with the equations of the shearing sheet, whereas (to the best of our knowledge) we are the only authors to have started with a global linear theory and then derived the shearing sheet results --- in particular the wavenumber `selection rule' encoded in the kernel \eqref{eqn:Volterra_Kernel_Sheet} --- as a consequence of well-defined assumptions. As a result, our \S\ref{sec:Small_Azimuthal} constitutes not only a rederivation but also a generalization of the work by \cite{julian1966non}.

An alternative view of spiral structure arising in the 1960s was that spirals are neutral modes of the background axisymmetric disk. This hypothesis of `quasi-stationary spiral structure' was promulgated especially by \cite{lin1966spiral,lin1970interpretation} and coworkers. In our language, the most basic form of this theory focused on small-$m$ spirals in the tightly wound limit (\S\ref{sec:Small_Radial}), and used the van Kampen representation (\S\ref{sec:van_Kampen}) to derive LSK modes (\S\ref{sec:LSK}). A significant problem with this was that van Kampen (and hence LSK) modes form a continuum in frequency space. Thus, although any individual mode is a stationary oscillation, \textit{packets} of these waves are not; \cite{toomre1969group} differentiated \eqref{eqn:LSK_Dispersion_Relation} and showed that the associated group velocity would drive wavepackets across a galaxy disk in much less than a Hubble time. Another issue with the Lin-Shu program  was the fact that the LSK dispersion relation \eqref{eqn:LSK_Dispersion_Relation} forbids waves near corotation, even though one expects them to be naturally amplified there (\S\ref{sec:swing_amplification}). Thus, while the LSK dispersion relation is informative, it is not, on its own, a viable candidate for a comprehensive theory of spiral structure in \textit{any} galaxy (let alone \textit{all} galaxies).

These issues led various authors to abandon purely local theories, and instead turn to the problem of global spiral structure. The global linear response problem was formulated by \cite{shu1970density1} in the Volterra representation (in a similar form to our \S\ref{sec:Linear}) and by \cite{kalnajs1971dynamics} in the Landau representation (culminating in the `dispersion relation' \eqref{eqn:Landau_Dispersion_Relation}).
Subsequently, there were several schools of thought about how the global problem ought to be tackled.

One approach was to embrace the generic global linear response problem (equations \eqref{eqn:total_potential_fluctuation}-\eqref{eq:linear_Vlasov_formal_solution}) without making any asymptotic wavelength approximations --- see, e.g., \cite{dekker1974rate,dekker1976spiral,Kalnajs1976-gg,kalnajs1977dynamics}. This has since led to much numerical `mode finding' (often using the Landau relation \eqref{eqn:Landau_Dispersion_Relation}), by, e.g., \cite{Evans1994-yc,evans1998stability1,evans1998stability2,Pichon2006-ak,jalali2005unstable,jalali2007unstable,jalali2008unstable,petersen2023predicting} and many others.

A second approach was to focus on --- in our language --- intermediate/long-wavelength spirals, effectively  reducing the number of $n_R$ components that need to be considered and highlighting the role of corotation and inner/outer Lindblad resonances (see the kernel \eqref{eqn:Volterra_Kernel_Long}). This was the approach taken, e.g., by \cite{Lynden-Bell1972-ve}, who used the Lagrangian representation to follow the angular momentum transferred to and from stellar orbits in spiral fields; in several places in the book by \cite{palmer1994stability}, who mostly used the Landau representation; and in papers by \cite{polyachenko2004outline,polyachenko2004unified}. Reassuringly for this approach, the global numerical calculations by \cite{jalali2005unstable} showed that --- at least for soft-centered disks --- the $n_R=0,\pm 1$ contributions truly do dominate the global Landau modes. In other words, such spirals should approximately obey the long-wavelength dispersion relation \eqref{eqn:Dispersion_Relation_Long}\footnote{An important caveat here is that \cite{jalali2005unstable} found significant contribution coming from near-radial orbits, which are not included in the epicyclic approximation underlying our work.}.

A third, hybrid, approach was also put forward in an attempt to rescue the quasi-stationary program of Lin and Shu. The key idea was to retain local LSK-like waves as the backbone of the spiral throughout most of the disk, but to acknowledge that they are an incomplete description near resonances (especially corotation), and that (given the group velocity problem and other effects like quasilinear damping) fresh waves need to be constantly generated. The natural language for this approach is a general WKB theory based on the formula \eqref{eqn:WKB_WLOG}. Simply put, one supposes that LSK is a good local description in quiescent parts of the disk, and then constructs global solutions by gluing together several such regions, dealing with special locations like resonances separately.
These special locations are just interfaces where LSK waves might be locally reflected/transmitted/absorbed/amplified. One can then interpret global neutral modes and/or instabilities as superpositions of counter-propagating wavetrains.
This idea lies behind a large amount of formal work, such as that by \cite{shu1970density2,Mark1974-jj,mark1976density,Mark1976-pz,Mark1976-xx,Mark1977-yx}; in particular, the influential paper by \cite{Mark1976-pz} introduced the `WASER' mechanism, in which a key role is played by the swing amplifier at corotation (\S\ref{sec:swing_amplification}). Closely related is the work on `quantum' conditions and dispersion relations for general WKB waves including open waves by, e.g., \cite{marochnik1969hypothesis,bertin1989modal}. See \cite{bertin1996spiral,griv2000landau,bertin2014dynamics,dobbs2014dawes} for further references.
 

Another important development  has been the search for spiral instabilities in disks with sharp features in the DF. This line of research was instigated by simulations of \cite{sellwood1989recurrent} and associated (Lagrangian) linear theory of \cite{sellwood1991spiral} (see also \citealt{lovelace1978negative, palmer1994stability}), the key analytic result of which we recovered in \S\ref{sec:Long_Wavelength_Instability_Sharp}. In particular, the groove instability (equation \eqref{eqn:Dispersion_noDehnen_CROnly_expanded}) plays a central role as the piece of linear physics upon which Sellwood \& Carlberg's (nonlinear) theory of spiral generation and maintenance is built (see \S\ref{sec:Nonlinear}). More recent numerical work has uncovered spiral instabilities related not only to grooves \citep{de2016spiral,De_Rijcke2019-uo} but also to disk breaks \citep{fiteni2024role}; all of these results should be (approximately) recoverable by implementing numerically the long-wavelength dispersion relation \eqref{eqn:Dispersion_Relation_Long}, although we have not verified this.


Finally, several authors have attempted to `bridge' or `unify' different aspects (and formulations) of spiral structure theory, especially the \cite{julian1966non,goldreich1965ii} swing amplifier on the one hand and the LSK dispersion relation \eqref{eqn:LSK_Dispersion_Relation}, or some generalization thereof, on the other. Examples include \cite{Goldreich1978-xe,drury1980normal,lin1987spectral,bertin1989modal, shu2000singular, yuan2001resonance, fuchs2001density, meidt2024bottom}\footnote{In particular, Appendix A of \cite{meidt2024bottom} includes a useful comparison between the many asymptotic dispersion relations derived in the short-wavelength, gaseous regime.}. However, the majority of these start from a local (e.g., shearing sheet) approach.
Those that start from a truly global approach have been exclusively fluid-dynamical (e.g., \citealt{shu2000singular}). It appears that nobody studying stellar disks has taken the approach of the present paper, namely to start with the global linear theory problem as in \S\ref{sec:Linear} and show that all more specialized linear theories arise as special cases. 

Perhaps the work that most closely resembles ours is \cite{fouvry2015secular1,fouvry2015secular2,fouvry2015secular3,fouvry2015self} (who used the Landau representation) and \cite{Dootson2022-cu,dootson2023bars} (in the Volterra representation). 
These authors embraced the global angle-action based approach and derived a lot of useful results, but in their fully-analytic work tended to fall back on the tight-winding approximation (\S\ref{sec:Small_Radial}).

\subsection{Nonlinear effects and open questions}
\label{sec:Nonlinear}

We have based our spiral structure theory around linear response (\S\ref{sec:Linear}) and ignored nonlinear effects. Although we gave a brief justification for this in \S\ref{sec:Simplifying_Assumptions}, it will certainly break down in real galaxies eventually. Here we identify (some) scenarios in which nonlinear effects are important and point out associated literature, as well as highlighting a few open questions.


On what timescale do we expect the linear approximation to break down? It depends on the amplitude and coherence time of the fluctuations in the gravitational potential (in other words, on their Kubo number). For instance, suppose the typical non-axisymmetric (spiral) component of the potential fluctuations is $\vert \delta \phi / V^2\vert\sim \eta$.
Then many short-lived, incoherent random spirals (i.e. a set of fluctuations with small Kubo number) will drive evolution of stellar orbits on a quasilinear timescale  \citep{Binney1988-zy}
\begin{equation}
    t_\mathrm{QL} \sim  \eta^{-2} t_\mathrm{dyn}
\end{equation}
Since $\eta \lesssim 0.1$, this quasilinear time will tend to be $t_\mathrm{QL} \gtrsim 10^2\,t_\mathrm{dyn}$, likely comparable to the age of the galaxy in question. Though this estimate is extremely rough, it gives us confidence that linear theory should hold in the small Kubo-number limit at least for a few Gyr. On the other hand, a single, coherent long-lived spiral (the large Kubo-number regime) will transport orbits nonlinearly on a timescale \citep{sellwood2002radial}
\begin{equation}
    t_\mathrm{lib} \sim \eta^{-1/2}t_\mathrm{dyn},
\end{equation} 
which may be only a handful of orbital periods (perhaps $\sim 1$ Gyr), for $\eta \lesssim 0.1$. Other nonlinear effects that we will discuss below all occur in the small Kubo-number regime and on a typical timescale $\gtrsim t_\mathrm{lib}$. Thus, we will say generally that linear theory is valid as long as $t\ll t_\mathrm{NL}$ where
\begin{equation}
    t_\mathrm{NL} = \mathrm{min}\left( t_\mathrm{QL}, t_\mathrm{lib} \right),
\end{equation}
while for $t\gtrsim t_\mathrm{NL}$ we must turn to nonlinear theory.

The most important nonlinear spiral mechanism is probably the `groove cycle' of Sellwood \& Carlberg. A sharp feature in the DF (\S\ref{sec:Long_Wavelength_Instability_Sharp}) provokes an instability, which then saturates nonlinearly (likely by trapping at corotation, see \citealt{sellwood2022spiral,hamilton2024saturation}), and decays by resonant (likely quasilinear) damping at its inner Lindblad resonance \citep{sellwood2019spiral}, thereby carving a new groove that itself is linearly unstable; and so the cycle repeats. This sequence of events explains spiral behavior in idealized global simulations \citep{sellwood2019spiral,Sellwood2020-ao}, and naturally produces few-armed grand-design spirals on realistic timescales (see \S\ref{sec:amplitude_requirements}). It is still an open question as to how often this mechanism occurs in real galactic disks with gas and star formation. On one hand, the formation of new stars keeps the disk cool enough to allow repeated, vigorous instabilities \citep{Sellwood1984-bw}, aiding the groove cycle. On the other hand, it is not clear how the first groove is produced (although see \citealt{Sellwood2012-ro,roule2025long} for idealized examples in which it is a natural outcome of secular evolution). In this paper we have ignored any `source' terms in our linearized kinetic equation \eqref{eq:Fluctuation_Evolution} that would result from new stars being born, as well as any local diffusion processes, although these could in principle be incorporated (as in the top box in Figure \ref{fig:Summary}).

Another influential nonlinear phenomenon is the `self-perpetuating' spiral wake behavior discovered by \cite{d2013self}. These authors placed massive perturbers on circular orbits in global simulations of a stellar disk, and showed that the perturbers initially produced \cite{julian1966non}-type wakes as anticipated by linear theory (\S\ref{sec:swing_amplification}), but that these wakes persisted long after the perturbers were removed. This lies beyond the short-wavelength linear analysis of \S\ref{sec:Short_Wavelengths}, and indeed there is currently no quantitative theory of this phenomenon.

For completeness we mention two nonlinear mechanisms that may occur in barred galaxies.
The first is nonlinear bar-spiral coupling \citep{Tagger1987-sf,Sygnet1988-eq,Masset1997}, the analytic theory for which recalls much of the long wavelength analysis we developed in Paper I and in \S\ref{sec:Long_Wavelengths}. The second is the `manifold theory' \citep{romero2007formation,athanassoula2012manifold}, which came out of the realization that in galaxies with strong bars, stellar orbits would be nonlinearly perturbed by the bar's non-axisymmetric gravitational field and that they would pile up at various points along an `invariant manifold' whose real-space imprint has a spiral structure. These two mechanisms are plausible drivers of few-armed spirals in strongly barred galaxies, although obviously do not fashion explanations for many-armed spirals or spirals in unbarred galaxies.

\section{Summary}
\label{sec:Summary}

In this paper, we have studied linear spiral structure in the stellar components of thin galactic disks. Our work can be summarized as follows.
\begin{itemize}
    \item We identified the characteristic scales in the spiral structure problem (\S\ref{sec:Characteristic_Scales}) and put quantitative requirements on the amplification factors $\mathcal{A}$ (\S\ref{sec:amplitude_requirements}) that must be achieved to match observations. Often $\mathcal{A}\sim 10^2$ or more (equation \eqref{eqn:amplitude_vs_observation}), placing stringent constraints on the portions of parameter space relevant for spiral structure theory.
    
    \item We formulated the general linear response theory for thin stellar disks (\S\ref{sec:General_Linear}), culminating in the Volterra equation \eqref{eqn:Volterra_Equation}, whose behavior we investigated numerically (\S\ref{sec:Linear_Theory_Example}). We could have developed the theory in an alternative representation (Landau, van Kampen, ...), although we argued that there are still some technical problems to be surmounted (\S\ref{sec:choice_of_representation} and Appendix \ref{sec:App_Representations}). We noted that when comparing two or more spiral ideas one should make sure they are truly physically distinct rather than just expressed in different vocabulary. We argued that self-gravity should be ignorable (equation \eqref{eqn:small_kernel}) in the regimes of large $Q$ (Appendix \ref{sec:large_Q}) and very short wavelengths.
    
    \item We investigated spirals at long wavelengths (\S\ref{sec:Long_Wavelengths}). We showed how to simplify the Volterra kernel in this regime (\S\ref{sec:Volterra_Long}). Then we showed that when self-gravity is omitted, long-wavelength spirals are often well-described as Lindblad-Kalnajs kinematic density waves (\S\ref{sec:Long_Wavelengths_Weak_Self_Gravity}), which are naturally two-armed (\S\ref{sec:Why_Two_Armed}); we also clarified the terminology surrounding density wave theory (Appendix \ref{sec:density_waves}). We further considered global spiral instabilities at long wavelengths (\S\ref{sec:Global_Instabilities}). We provided a new derivation of the dispersion relation for the groove instability (\S\ref{sec:Long_Wavelength_Instability_Sharp}) and used this to argue that the maximum growth rate of spiral instabilities is of order the dynamical frequency (\S\ref{sec:max_growth_rate}).
    
    \item We considered the limit of short wavelengths (\S\ref{sec:Short_Wavelengths}). We simplified the Volterra kernel in this regime (\S\ref{sec:Volterra_short}) and used the result to confirm the weakness of self-gravity at the smallest scales (Figure \ref{fig:Volterra_Kernel_Illustration_Wavelength_Dependence}). We then considered two limits of short-wavelength theory: first the limit of short azimuthal wavelengths (large $m$, \S\ref{sec:Small_Azimuthal}), in which we recovered (and generalized) the classic shearing sheet results of \cite{julian1966non}, and second the tightly wound limit (small $m$ \textit{and} small radial wavelengths, \S\ref{sec:Small_Radial}), wherein we recovered the LSK dispersion relation (\S\ref{sec:LSK}). We found that the tightly wound Volterra kernel connects smoothly to the low-$m$ extrapolation of the large-$m$ kernel from \S\ref{sec:Small_Azimuthal}.
    
    \item We provided an extensive discussion of previous (linear) theories of spiral structure and how they fit into --- or can be derived from --- the picture developed here (\S\ref{sec:previous_literature}). We also mentioned some important nonlinear phenomena that are likely to affect spiral behavior (\S\ref{sec:Nonlinear}).
    These are strictly beyond the scope of this paper, but are closely linked to linear results (Figure \ref{fig:Summary}).    
\end{itemize}

Let us emphasize that, similar to our previous \textit{galactokinetics} paper, much of what we have concluded here can already be found scattered across the vast literature on linear spiral structure that has gathered over the last six decades (\S\ref{sec:previous_literature}). But we now have for the first time a coherent framework in which the most important results of (analytic) linear theory can be derived  --- and indeed generalized  --- in a rigorous and systematic way.
And, not only do our reduced descriptions give us insight in particular asymptotic regimes, they also tend to connect smoothly when extrapolated beyond those regimes, giving us confidence that we are actually gaining insight into real spiral phenomena. With this work done, the most important unanswered questions in spiral structure theory are primarily \textit{nonlinear}. Combined with recent progress in numerical simulations of spiral behavior in the nonlinear regime \citep{sellwood2019spiral,roule2025long}, we hope that this paper provides a solid analytic foundation upon which a comprehensive theory of spiral structure might ultimately be built.


\begin{acknowledgments}
We are grateful to the many colleagues whose insights on this topic have improved our work, including U. Banik, J. Binney, J. Bland-Hawthorn, A. Burrows, K. Daniel, S. de Rijcke, W. Dehnen, C. Fairbairn, J.-B. Fouvry, E. Griv, J. Iga, M. Kunz, N. Magnan, J. Magorrian, S. Meidt, G. Ogilvie, E. Ostriker, M. Pessah, M. Petersen, C. Pichon, R. Rafikov, M. Roule, J. Sellwood, J. Stone, A. van der Wel, M. Weinberg, and T. Yavetz. 

C.H. is supported by the John N. Bahcall Fellowship Fund and the Sivian Fund at the Institute for Advanced Study. S.M. acknowledges support from the National Science Foundation Graduate Research Fellowship under Grant No. DGE-2039656.
\end{acknowledgments}


%


\appendix

\section{Alternative representations of linear theory}
\label{sec:App_Representations}

In this Appendix, we give some details of three approaches to linear response theory that differ from the Volterra approach we adopted in \S\ref{sec:Linear}. 

\subsection{Landau representation}
\label{sec:Landau}

One popular approach is to Laplace transform \eqref{eq:linear_Vlasov_formal_solution} and \eqref{eqn:B_formal} in time (or, equivalently, one could Laplace transform \eqref{eqn:Volterra_Equation} directly).
The objects of interest are then the potential and DF fluctuations in the space of \textit{complex} frequencies $\omega$. Having solved for these fluctuations one can, at least in principle, revert them to the time domain via inverse Laplace transform.
This was the approach originally taken by \cite{Landau1946-aj} to the analogous problem of self-consistent linear oscillations in collisionless electrostatic plasma, and so we will call it the \textit{Landau} representation.

Using the Landau approach, one expects to be able to write the solution for $\delta \phi_{\bn}(\bJ, t)$  in the form
\begin{align}
    &\delta \phi_{{\bm{n}}}(\bm{J}, t) =  \sum_{{\bm{n}}'}\int  \md \bm{J}' \, a_{{\bm{n}}{\bm{n}}'}(\bm{J},\bm{J}') \, \me^{ - \mi {\bm{n}}' \cdot \bOm't}+ \sum_g b^ g_{\bn \bm{n}'}(\bm{J},\bm{J}') \me^{- \mi \omega_g t}.
    \label{eqn:dressed_potential_fluctuations}
\end{align}
Here $a, b$ are complicated functions of the initial conditions and the external forcing, and $\omega_g$ are the frequencies of any \textit{Landau modes},
which are solutions of
\begin{equation}
    \mathrm{det} [\mathcal{I}-\widehat{\mathcal{M}}(\omega)] = 0,
    \label{eqn:Landau_Dispersion_Relation}
\end{equation}
where the \textit{response matrix} $\widehat{\mathcal{M}}$ is the Laplace transform of the Volterra kernel \eqref{eqn:Volterra_Kernel}:
\begin{align}
\widehat{\mathcal{M}}^{pp'} (\omega) 
&= \int_0^\infty \md t\,\me^{i\omega t}\mathcal{M}^{pp'}(t) \label{eqn:Laplace}
\\&= \frac{(2 \pi)^2}{\mcE} \sum_{{\bm{n}}} \!\! \int \!\! \md \bm{J} \, \frac{{\bm{n}} \cdot \p f_{0} / \p \bm{J}}{\omega - {\bm{n}} \cdot \bOm } 
    [\phi^{(p)}_{\bm{n}} (\bm{J})]^*
    \phi^{(p')}_{\bm{n}} (\bm{J}).
\label{Fourier_M}
\end{align}
(The $\bJ$ integration has to be defined via analytic continuation in the case Im\,$\omega < 0$.) The Landau approach has the advantage that the time-dependence in solutions like \eqref{eqn:dressed_potential_fluctuations} is, formally at least, explicit. It also makes manifest the key role of any exponentially-growing instabilities, which will dominate the solution \eqref{eqn:dressed_potential_fluctuations} after sufficient time. In the context of spiral structure theory, the Landau representation  has been used for the development of formal theory by, e.g., \cite{kalnajs1971dynamics,dekker1976spiral}, and for numerical calculations (see, e.g., \cite{petersen2023predicting} and references therein).

However, there are several issues with the Landau approach.  First, it can be tricky to implement numerically, especially if there are important damped Landau modes (Im\,$\omega_g < 0$). Second, in deriving a solution of the form \eqref{eqn:dressed_potential_fluctuations},  one has made certain assumptions about the smoothness and analyticity of the DF and the action space domain; relaxing these assumptions would lead to extra pieces of the solution that decay non-exponentially, as has been demonstrated for the Hamiltonian Mean Field model by \cite{Barre2013}.

Third, a more conceptual (but possibly also more serious) issue is that, as far as we know, the correct angle-action linear response calculation including Landau modes --- i.e. an equation of the form \eqref{eqn:dressed_potential_fluctuations} with explicit $a$ and $b$ --- has never actually been written down in stellar dynamics. The reason is that in order to get the explicit `sum over Landau modes' part of the solution in \eqref{eqn:dressed_potential_fluctuations}, formally one has to assume those Landau modes are mutually orthogonal. This is not guaranteed since the linear operator equation whose solution we are seeking is not hermitian\footnote{It is also non-hermitian in homogeneous plasma, but this is not an issue since one usually considers a single mode at each $\bk$, and these \textit{are} orthogonal --- see \cite{Rogister1968-tb}.}. One can approximate the operator as nearly hermitian if modes are weakly damped \citep{Layzer1963-ml}; after some additional assumptions about the operator's behavior near Landau modes \citep{Kaufman1970-za,Kaufman1971-ld,Kaufman1972-vv} one arrives at the form \eqref{eqn:dressed_potential_fluctuations} with explicit (but by no means easy to calculate) functions $a$ and $b$. However, the details of this argument in the stellar-dynamical case have never been published, nor have its assumptions been tested numerically (but see El Rhirhayi et al., in prep.)

Thus, we feel that the Landau approach to linear theory is incompletely understood at present. It is most useful when the system harbors an unstable Landau mode: then one can use the Landau approach to search for the instability, knowing that it will dominate the linear response at late times, and not worry about the above subtleties.


\subsection{van Kampen representation}
\label{sec:van_Kampen}

Another approach is to simply look for solutions of equations \eqref{eq:linear_Vlasov_formal_solution} and \eqref{eqn:B_formal} 
for which $\delta f_{\bn}(\bJ,t)$ and $\delta \phi_{\bn}(\bJ,t)$ have time dependence $\propto \exp(i\omega  t)$, for some fixed (possibly complex) $\omega$.  In the case where $\omega$ is real these are `normal modes' of the system in the traditional sense of these words. More generally, the set of solutions found by this method are called `van Kampen modes', and so we will call this  the \textit{van Kampen representation}. In the van Kampen representation one finds that there is in fact a continuum of solutions $\propto \exp(-i\omega  t)$ along the real $\omega$ axis. In addition, if the system is unstable, i.e., if there exist any exponentially growing Landau mode according to the Landau representation (some $\omega_g$ such that det$[\mathcal{I}-\mathcal{M}(\omega_g)]$ and Im\,$\omega_g > 0$), then one must supplement this continuum of solutions with extra solutions proportional to both $\exp(-i\omega_g  t)$ and $\exp(-i\omega_g^* t)$.


An advantage of the van Kampen representation is that, in the case of stable systems, it avoids complex frequencies entirely.  However, there are several disadvantages. First, unless the system is particularly symmetric, one often has to perform a Landau calculation anyway in order to \textit{know} the system is stable. Second, the real frequenciy solutions are actually singular distributions rather than true functions, i.e., they only have physical meaning when integrated over some patch of phase space. Third, the mathematical theory of van Kampen modes in generic stellar systems --- particularly in rotating systems like disks --- is not yet fully mature. Importantly, it has not been shown that these modes form a complete set of orthogonal functions in terms of which an arbitrary solution to the initial value problem can be constructed. The current state of the art on this topic can be found in \cite{Lau2021-uf,lau2021modes} (see also \citealt{ng2021landau}).

\subsection{Lagrangian representation}
\label{sec:Lagrangian}

All three of the methods just mentioned are Eulerian in the sense that $(\br, \bm{v})$ or $(\bm{\theta}, \bJ)$ are just coordinates on phase space, and the quantities one calculates are the values of certain fields (potential, DF, etc.)  at a given phase space location and time $t$. A fourth alternative is a Lagrangian method in which one actually solves for the evolution of stars' phase-space coordinates $\bm{\theta}(t)$ and $\bJ(t)$ given their initial conditions $\bm{\theta}(0)$ and $\bJ(0)$. We will call this the \textit{Lagrangian representation}. This approach was used in the spiral context by, e.g., \cite{Lynden-Bell1972-ve}, and by \cite{toomre1981amplifies} in his investigation of swing amplification (but see \S\ref{sec:previous_literature}). A more recent example of the Lagrangian formalism at work in a slightly different context is \citealt{binney2024disc}.


\section{The weakness of self-gravity for large $Q$}
\label{sec:large_Q}

We expect that \eqref{eqn:small_kernel} is true at all wavelengths, provided we make $Q$ large enough. To see that this is the case, note that the only place $Q$ explicitly enters the general Volterra kernel \eqref{eqn:Volterra_Kernel} is via $f_0$.  We can now imagine varying the three parameters involved in $Q$ --- namely $\sigma, \Sigma$ and $\kappa$ --- one at a time while keeping the other two fixed. We will consider how \eqref{eqn:Volterra_Kernel} scales with $Q$ in each of these cases. We suppose for concreteness that $f_0$ is given by the Schwarzschild distribution \eqref{eqn:Schwarzchild}, so the total mass in stars is
\begin{equation}
    M = (2\pi)^2\int \md \bJ f_0 = (2\pi)^2 \int \md J_\varphi \langle J_R \rangle \, g_0.
    \label{eqn:mass_in_disk}
\end{equation}
In particular, for a fixed surface density profile we expect $g_0 \propto 1/\langle J_R\rangle$.\footnote{Strictly speaking, $\md M/\md J_\varphi = -\Sigma(R_\mathrm{g})/(2B)$ where $B$ is the Oort constant \eqref{eqn:Oort_Constants}. Thus, $g_0 \propto 1/\langle J_R\rangle$ really corresponds to a fixed profile of $\Sigma/B$ rather than $\Sigma$. But our arguments here are all local in $J_\varphi$ so this distinction is not important and we will refer to `fixed surface density' throughout.}

First, if we fix $\sigma$ and $\kappa$ but vary $\Sigma$ (e.g., by changing the mass of the stars in our disk), then $\langle J_R \rangle = \sigma^2 /\kappa$ is independent  of $Q$ while $g_0 \propto \Sigma\propto 1/Q$. Thus
\begin{equation}
    \frac{\p f_0}{\p J_\varphi} \propto \frac{\p f_0}{\p J_R} \propto \frac{1}{Q},
\end{equation}
and so $\mathcal{M}^{pp'}\propto 1/Q \to 0$ as $Q\to \infty$. It makes obvious physical sense  that by  decreasing the mass of our disk we will diminish the role of self-gravity and cause our stars to act like test particles.

Second, if we fix $\Sigma$ and $\kappa$ but vary $\sigma$ (e.g., by initializing our stars with the same distribution of guiding radii but different eccentricities), then $\langle J_R \rangle \propto Q^2$ and  $g_0 \propto 1/\langle J_R\rangle \propto 1/Q^2$.  Thus, schematically,
\begin{equation}
    \frac{\p f_0}{\p J_\varphi} \propto \frac{1}{Q^2} \me^{-J_R/Q^2},\,\,\,\,\,\,\,\,\,\,\,\, \frac{\p f_0}{\p J_R} \propto \frac{1}{Q^4} \me^{-J_R/Q^2}.
\end{equation}
In this case all terms involving $\p f_0/\p J_R$ go to zero at large $Q$. Physically this is because by increasing $\sigma$ we are spreading out the population of stars in frequency space, so that the phase $\me^{-i\bn\cdot\bOm t}$  evolves very differently across the population even at a fixed $J_\varphi$, causing the integrand in the Volterra kernel \eqref{eqn:Volterra_Kernel} to oscillate rapidly in $\bJ$ at a fixed $t$. In the epicyclic approximation this manifests in the fact that almost all terms in our asymptotic expressions \eqref{eqn:Volterra_Kernel_Long} and \eqref{eqn:Volterra_Kernel_Short} tend to zero as the Dehnen drift $\zeta \propto Q^{2} \to \infty$.

Third, we imagine that we fix $\sigma$ and $\Sigma$ but change $\kappa$. This is a rather artificial prospect, since there is no way to change $\kappa$ significantly without changing $\Omega$, which would change the angle-action mapping, which changes everything else in the Volterra kernel calculation, too. But let us proceed regardless; we have $\langle J_R \rangle \propto 1/Q$ while  $g_0\propto 1/\langle J_R\rangle \propto Q$, and so
\begin{equation}
    \frac{\p f_0}{\p J_\varphi} \propto Q\,\me^{-QJ_R},\,\,\,\,\,\,\,\,\,\,\,\, \frac{\p f_0}{\p J_R} \propto Q^2\,\me^{-QJ_R}.
\end{equation}
In this case, contributions to $\mathcal{M}^{pp'}$ from finite $J_R$ go to zero as $Q\to \infty$ --- physically, we have created a situation in which all interactions that involve radial phases oscillate extremely rapidly. On the other hand, contributions to $\mathcal{M}^{pp'}$ from  essentially circular orbits ($J_R\to 0$) do not vanish as $Q\to \infty$, because circular orbits are insensitive to the radial frequency profile.


We conclude that {increasing $Q$} is \textit{nearly} a sufficient condition for killing self-gravitating effects in linear spiral structure theory. The only exception arises in the (artificial) third case just discussed.



\section{Volterra kernel in asymptotic wavelength regimes}

In this Appendix we derive expressions for the Volterra kernel \eqref{eqn:Volterra_Kernel_l} in the long (\S\ref{sec:Long_wavelength_Kernel}) and short (\S\ref{sec:WKB_Kernel}) asymptotic wavelength regimes, using the approximate results we developed in Paper I (which ignore corrections $\mathcal{O}(\epsilon_R^2)$).

However, we must first recall a technical detail from Paper I, which is that in some circumstances we had to include the  `Dehnen drift' correction to the epicyclic approximation. Though formally $\mathcal{O}(\epsilon_R^2)$, this correction was necessary to make the mapping from $(\br,\bm{v})\to(\btheta,\bJ)$ canonical. Empirically, we found it was usually negligible when calculating the Fourier coefficients of the potential fluctuations, but was \textit{not} negligible when performing a quasilinear transport calculation. The reason was that the latter calculation involved an additional term $\propto \me^{-in_\varphi \Omega_\varphi t}$, and the Dehnen drift correction to $\Omega_\varphi$ (equation (52) of Paper I) caused a gradual phase offset which was crucial near resonances.  Following this lesson, when simplifying the Volterra kernel \eqref{eqn:Volterra_Kernel_l} we shall ignore the Dehnen drift correction to the functions $u_{\ell n_R}^{q}$ and $u_{\ell n_R}^{q'}$, but we will include the correction to the factor $  \me^{-in_\varphi\Omega_\varphi \tau}$ that multiplies them.

\subsection{Long-wavelength regime}
\label{sec:Long_wavelength_Kernel}

From equation (73) of Paper I, at long wavelengths we have 
\begin{align}
         u^{q}_{\bn}(\bJ) = \delta_{n_\varphi}^\ell u^q(\Rg)
     \times  \begin{cases}
        1, & n_R=0, \\ 
        \frac{\ell \gamma a_R}{2 R_\mathrm{g}} \left[ \pm 1- \frac{i k^q_R R_\mathrm{g}}{\ell \gamma}\right], & n_R = \pm 1,
        \\
        0, & \vert n_R\vert \geq 2,
    \end{cases} 
\end{align}
where $k^q_R$ is the radial wavenumber of the basis function $u^q$ (equation \eqref{eqn:Radial_Wavenumber_q}). The sum over $n_R$ in \eqref{eqn:Volterra_Kernel_l} can then be carried out explicitly, with the result 
\begin{align}
    \mathcal{M}^{\ell qq'}(\tau) &= 
    -  \frac{(2\pi)^2i }{\mathcal{E}} 
    \int \md J_\varphi [u^{q}(\Rg)]^*u^{q'}(\Rg)
    \int_0^\infty \md J_R \, \Bigg\{ 
    \ell \frac{\p f_0}{\p J_\varphi}\me^{-i\ell\Omega_\varphi\tau} 
    \nn
    \\
    &\,\,\,\,\,\,\,\,\,\,\,\,\,\,\,\,\,\,\,\,\,\,\,+ \left( \ell \frac{\p f_0}{\p J_\varphi} + \frac{\p f_0}{\p J_R} \right) \xi_+^{\ell qq'}(\Rg,a_R)
    \me^{-i(\ell\Omega_\varphi +\kappa) \tau}
    + \left( \ell \frac{\p f_0}{\p J_\varphi} - \frac{\p f_0}{\p J_R} \right)
    \xi_-^{\ell qq'}(\Rg, a_R)
    \me^{-i(\ell\Omega_\varphi -\kappa) \tau}
    \Bigg\},
    \label{eqn:Long_Volterra_Step_1}
\end{align} 
where we defined the dimensionless functions
\begin{equation}
    \xi_{\pm}^{\ell qq'}(\Rg, a_R) \equiv \left(\frac{\ell \gamma a_R}{2 R_\mathrm{g}} \right)^2 \bigg[\pm 1 - \frac{ik^q_R R_\mathrm{g}}{\ell \gamma}\bigg]^* \bigg[ \pm 1 - \frac{ik_R^{q'} R_\mathrm{g}}{\ell \gamma}\bigg].
    \label{eqn:dimensionless_functions}
\end{equation}

If we now assume a Schwarzschild DF \eqref{eqn:Schwarzchild}, and apply the identity
\begin{align}
    &\int_0^\infty \md x \,x^n \me^{-(1+ib)x}  = \frac{n!}{(1+ib)^{n+1}},
\end{align}
for $n=0,1,2$, then we can carry out the $J_R$ integral in \eqref{eqn:Long_Volterra_Step_1}, resulting in the expression \eqref{eqn:Volterra_Kernel_Long} given in the main text, with the auxiliary functions $G_0^{\ell}$ and $G_\pm^{\ell q q'}$ given by
\begin{align}
    &    G_0^{\ell}(J_\varphi, \tau) = 
    -  (2\pi)^2i \ell  g_0 \langle J_R \rangle  \bigg[ \frac{1}{\zeta} \frac{\md \ln g_0}{\md J_\varphi}  +
    \frac{1}{\zeta^2}  \frac{\md \ln \langle J_R \rangle}{\md J_\varphi}\bigg] 
    \label{eqn:aux_0}
    \\
    &    G_\pm^{\ell qq'}(J_\varphi, \tau) = 
    -  (2\pi)^2i \ell g_0 \langle J_R \rangle
    \bigg[\frac{1 }{\zeta^2}
    \frac{\md \ln g_0}{\md J_\varphi} 
    +
    \frac{2  }{\zeta^3}
    \frac{\md \ln \langle J_R \rangle}{\md J_\varphi}
    \mp
    \frac{1}{\ell \zeta^2 \langle J_R \rangle}
    \bigg]
    \xi_\pm^{\ell qq'}(R_\mathrm{g},a).
    \label{eqn:aux_pm}
\end{align} 
Note that the functions $\xi^{\ell qq'}_{\pm}$ in \eqref{eqn:aux_pm} are evaluated at $a_R=a$.

\subsection{Short-wavelength regime}
\label{sec:WKB_Kernel}

From Appendix B of Paper I we know that at short wavelengths, the basis elements \eqref{eqn:WKB_basis} have Fourier components
\begin{align}
    u^{q}_{\bn}(\bJ) &= 
    u(k_R) \me^{i(k_R R_\mathrm{g} + n_R \beta)}  (-i \, \mathrm{sgn}\,  k_R)^{n_R} J_{n_R}\left(K_R a_R\right),
\end{align}
where
\begin{equation}
    K_R \equiv \sqrt{k_R^2 + \frac{n_\varphi^2\gamma^2}{R_\mathrm{g}^2}} = \vert k_R \vert \sqrt{1 + \gamma^2 \bigg\vert \frac{k_\varphi}{k_R} \bigg\vert^2},\,\,\,\,\,\,\,\,\,\,\,\,\,\,\,\,\,\,\,\,\,
    \label{eqn:modified_k}
    \beta \equiv \arctan\left( \frac{ n_\varphi  \gamma}{ k_R  R_\mathrm{g}} \right) = \arctan\left( \frac{\gamma k_\varphi}{k_R}  \right),
\end{equation}
both depend on $k_R$, $n_\varphi$ and $J_\varphi$, but not on $J_R$. 
When we plug these coefficients into the Volterra kernel \eqref{eqn:Volterra_Kernel_l}, the sum over $n_R$ can be carried out using the following identities which follow from Graf's addition theorem \citep{abramowitz1965handbook}:
\begin{align}
    &\sum_{n=-\infty}^\infty J_n(x)J_n(y)\me^{-in\theta} = J_0\left( 
    \sqrt{x^2+y^2-2xy\cos\theta} \right),
    \\
        &\sum_{n=-\infty}^\infty J_n(x)J_n(y)\,n\,\me^{-in\theta} = -i \frac{xy\sin\theta}{    \sqrt{x^2+y^2-2xy\cos\theta}} J_1\left( 
    \sqrt{x^2+y^2-2xy\cos\theta} \right).
\end{align}
The result is $\mathcal{M}^{\ell q q'}(\tau)= \mathcal{M}^{\ell}(k_R,k_R',\tau)$ where
\begin{align}
    \mathcal{M}^{\ell}(k_R,k_R',\tau) = 
    -  \frac{(2\pi)^2i }{\mathcal{E}} 
    &\int \md J_\varphi [u(k_R)]^* u(k'_R) 
    \nn
    \\
    & \times
    \int_0^\infty \md J_R \,\me^{-i(k_R-k'_R)\Rg-i\ell\Omega_\varphi\tau}
    \Bigg[ 
    \ell\frac{\p f_0}{\p J_\varphi} J_0(\mathcal{K}a_R)- \frac{\p f_0}{\p J_R}
    \frac{i K_RK_R' a_R\sin\theta
    }{\mathcal{K}} J_1(\mathcal{K} a_R)
    \Bigg].
    \label{eqn:almost_short_kernel}
\end{align}
Here we defined the `wavenumber'
\begin{equation}
    \mathcal{K}(k_R,k_R',\tau) = \sqrt{K_R^2+K_R'^2-2K_RK_R'\cos\theta},
    \label{eqn:calK}
\end{equation}
and the angle
\begin{equation}
    \theta(k_R,k_R',\tau) = \begin{cases}
        \kappa\tau + \beta - \beta', \,\,\,\,\,\,\,\,\,\,\,\,\,\,\,\,\,\,\,\,\,\, \mathrm{if} \,\,\,\,\,\,\,\,\,\,\, k_Rk_R' > 0
        \\
        \kappa\tau + \beta - \beta' - \pi, \,\,\,\,\,\,\,\,\,\,\, \mathrm{if} \,\,\,\,\,\,\,\,\,\,\, k_Rk_R' < 0,
    \end{cases}
    \label{eqn:time_dependent_angle}
\end{equation}
and $K_R'$, $\beta'$ are the same as $K_R$, $\beta$ (equation \eqref{eqn:modified_k}) except with $k_R\to k_R'$, and we set $n_\varphi = \ell$ in all their definitions.

If we further assume that $f_0$ has the Schwarzschild form \eqref{eqn:Schwarzchild}, then we can perform the integral over $J_R$ in \eqref{eqn:almost_short_kernel} explicitly. Using  the identities
\begin{align}
    &\int_0^\infty \md x \,x\,  \me^{-(1+ib)x^2 }J_0(cx) =    \frac{1}{2(1+ib)}\me^{-c^2/[4(1+ib)]},
    \\
    &\int_0^\infty \md x \,x^3\,  \me^{-(1+ib)x^2 }J_0(cx) =    \frac{1+ib-c^2/4}{2(1+ib)^3}\me^{-c^2/[4(1+ib)]},
    \\
        &\int_0^\infty \md x \,x^2\,  \me^{-(1+ib)x^2 }J_1(cx) =    \frac{c}{4(1+ib)^2}\me^{-c^2/[4(1+ib)]},
\end{align}
we arrive at the result \eqref{eqn:Volterra_Kernel_Short}, where the auxiliary function $G^{\ell}(k_R,k_R',\tau)$ is defined as 
\begin{align}
    &    G^\ell(k_R,k_R',\tau) = 
    -   (2\pi)^2i  \ell  g_0\langle J_R \rangle 
    \bigg[ \frac{1}{\zeta} \frac{\md \ln g_0}{\md J_\varphi} +   \frac{1-(\mathcal{K}a)^2/(4\zeta)}{\zeta^2} \frac{\md \ln \langle J_R \rangle}{\md J_\varphi} + \frac{iK_RK_R'  \sin \theta}{\zeta^2 \ell\kappa}
    \bigg]  \me^{-i(k_R-k_R')\Rg}\me^{-(\mathcal{K}a)^2/(4\zeta)}.
    \label{eqn:aux_G}
\end{align}

\section{Basis elements in the short wavelength regime}
\label{sec:WKB_basis_elements}

The potential basis element in this regime is $\phi^{(p)}(\br) = u(k_R)\,\me^{i(\ell\varphi + k_RR)}$ (see equations \eqref{eqn:general_basis} and \eqref{eqn:WKB_basis}). The corresponding surface density basis element is, up to corrections $\mathcal{O}(\vert kR_*\vert^{-1})\sim\epsilon^2$,
\begin{equation}
    \Sigma^{(p)}(\br) = - \frac{k^{(p)}}{2\pi G} u(k_R)\,\me^{i(\ell\varphi + k_RR)},
    \label{eqn:Sigma_p}
\end{equation}
where $k^{(p)} = \vert \bk^{(p)}\vert = \sqrt{(k_\varphi^\ell)^2+k_R^2}$. Given \eqref{eqn:Sigma_p}, we can expand an arbitrary short-wavelength surface density fluctuation by analogy with \eqref{eqn:potential_fluctuation_WKB_expansion} as:
\begin{equation}
    \delta\Sigma(\br, t)   =  - \frac{1}{2\pi G} \sum_{\ell} \fint \md k_R \,B^\ell(k_R, t)\sqrt{(k_\varphi^\ell)^2 + k_R^2} \,u(k_R)\,\me^{i(\ell\varphi+ k_RR)}.
    \label{eqn:density_fluctuation_WKB_expansion}
\end{equation}

However, we have not yet specified the function $u(k_R)$. We now do so using the biorthogonality condition \eqref{def_basis_2}. We have 
\begin{align}
    \int  \md \br \, [\phi^{(p)}(\br)]^* \, \Sigma^{(p')} (\br) &= -\frac{k^{(p')}}{2\pi G} [u(k_R)]^*u(k_R') \times 2\pi \delta_\ell^{\ell'}\times \int_0^\infty \md R\,R\,\me^{-i(k_R-k_R')}R
    \nn
    \\
    &\simeq -\frac{2\pi k^{(p)} \vert u(k_R)\vert^2 R_*}{G} \delta_\ell^{\ell'}\delta(k_R-k_R'),
    \label{eqn:biorthogonality_WKB}
\end{align}
where to get the second line we used the expansion \eqref{eqn:expanding_R_and_J} and the fact that the basis functions are localized\footnote{\cite{fouvry2015secular1} performed a similar calculation, and gave a more careful justification for localized basis elements.} around $R_*$. Given \eqref{def_basis_2} and the definition \eqref{eqn:sum_to_integral}, we would like the right hand side of \eqref{eqn:biorthogonality_WKB} to equal $-\mathcal{E} \times \delta_{\ell}^{\ell'} \times 2\pi L^{-1}\,\delta(k_R-k_R')$, where $\mathcal{E}$ has units of energy and is independent of the choice of basis element. We can achieve this by choosing 
\begin{eqnarray}
    u(k_R) = \frac{u_0}{\sqrt{2\pi\,k^{(p)}R_*}} \,\,\,\,\,\,\,\,\,\,\,\,\,\implies \,\,\,\,\,\,\,\,\,\,\,\,\, \mathcal{E}= \frac{L}{2\pi} \frac{\vert u_0\vert ^2}{G} = \frac{L}{2\pi} \frac{2\pi \,k^{(p)}R_*\vert u(k_R)\vert^2}{G},
    \label{eqn:curlyE_WKB}
\end{eqnarray}
where $u_0$ is an arbitrary number with the same units as $u$, namely (velocity)$^2$. 

\section{A note on `density wave' terminology}
\label{sec:density_waves}

We give the name \textit{density wave} to any spiral-like disturbance $\delta f$ whatsoever, regardless of whether it is self-gravitating, subject to external forces, linear or nonlinear, etc. Most stars involved in such a disturbance at a fixed time do not move with the spiral pattern as time proceeds, i.e. the stars that sit at the very trough of a density wave today (satisfying equation \eqref{eqn:R_of_phi}) were typically not in the trough yesterday and will not be there tomorrow. The term `density wave' therefore encompasses almost all of modern spiral structure theory --- to say that `spirals are density waves' is just to say that `spirals are over/underdensities in the stellar DF'. (An analogous definition would hold for spirals in gaseous disks).  

Following \cite{Binney2008-ou}, we define a \textit{kinematic} density wave as any disturbance in the stellar DF that evolves purely kinematically (i.e. follows equation \eqref{eq:linear_Vlasov_formal_solution} with $\delta \phi^\mathrm{tot}=0$). In other words, these are over/underdensities that exist in a disk of stars that simply move on their mean-field orbits in the axisymmetric  potential of the galaxy. The Lindblad-Kalnajs waves of \S\ref{sec:Long_Wavelengths_Weak_Self_Gravity} are a special case.

The only `spirals' we have mentioned so far that would \textit{not} be categorized as density waves are the passive `material arms' discussed in \S\ref{sec:Characteristic_Scales}. These consist of arbitrary tracers (like stripes of paint) on circular orbits in the galaxy's axisymmetric potential, and need not have anything to do with overdensities of stars/gas.

\section{Recovering the JT equation}
\label{sec:Recover_Binney}

\cite{binney2020shearing} derived a Volterra equation for a shearing component of the surface density fluctuations, starting from the equations of the shearing sheet. He called this the JT equation, after \cite{julian1966non} who first derived it. In this Appendix we recover (Binney's version of) the JT equation using the results from \S\ref{sec:Small_Azimuthal}. We proceed in two stages. First, we transform the results of Appendix \ref{sec:WKB_basis_elements} to shearing sheet coordinates, and hence write down a formal expression for a swinging surface density fluctuation (\S\ref{sec:coord_transform}). Second, we incorporate the time-dependence of this fluctuation using the Volterra equation (\S\ref{sec:JT_equation}).

\subsection{Coordinate transform}
\label{sec:coord_transform}


We consider the general surface density expansion \eqref{eqn:density_fluctuation_WKB_expansion}. Following \cite{binney2020shearing}, we now make three coordinate changes. First, Binney took his origin of time to be at $t_\mathrm{i}$ rather than zero. We achieve this by replacing  the lower integration limit in \eqref{eqn:Volterra_Equation_Short} with $t_\mathrm{i}$, and replacing $\delta f_{\bn}(\bJ, 0)\,\me^{-i\bn\cdot\bOm t} \to \delta f_{\bn}(\bJ, t_\mathrm{i})\,\me^{-i\bn\cdot\bOm (t-t_\mathrm{i})} $ in the definition of $B^\ell_\mathrm{kin}(k_R, t)$,   equation \eqref{eqn:Bkin_WKB}. Second, Binney employed shearing sheet coordinates $(x, y)$, which are related to our $(\varphi, R)$ by
\begin{eqnarray}
    x = R-R_*,
    \,\,\,\,\,\,\,\,\,\,\,\,\,\,\,\,\,\,\,\,\,\,\,\,\,\,\,\,\,\,\,\,\,\,\,\,
    y = R_*[\varphi - \Omega(t-t_\mathrm{i})].
           \label{eqn:Binney_Coordinates}
\end{eqnarray}
Third, Binney defined the wavenumber
\begin{equation}
    k_y = k_\varphi^\ell = \frac{\ell}{R_*}.
    \label{eqn:Binney_Wavenumbers}
\end{equation}
Plugging the transformations \eqref{eqn:Binney_Coordinates}-\eqref{eqn:Binney_Wavenumbers} into the right hand side of \eqref{eqn:density_fluctuation_WKB_expansion}, we find
\begin{equation}
    \delta \Sigma(\br, t) = \sum_{k_y} \fint \md k_R {\Sigma}_1(k_R, k_y, t)\,\me^{i(k_R x+ k_yy)},
\end{equation}
where the time-dependent coefficients are
\begin{equation}
    {\Sigma}_1(k_R, k_y, t) \equiv -\frac{u(k_R)}{2\pi G}\sqrt{k_R^2 +k_y^2} \,\,B^\ell (k_R, t) \,
    \me^{i[ \ell\Omega(t-t_\mathrm{i})+k_RR ]}.
    \label{eqn:Sigma_1_complicated}
\end{equation}
and the entire right hand side should be evaluated at $R=R_*$, although we drop the asterisks from now on to avoid cluttering the notation. The function ${\Sigma}_1$ is the primary object of the JT equation.

\subsection{Time evolution}
\label{sec:JT_equation}

The right hand side of \eqref{eqn:Sigma_1_complicated} evolves in time according to the Volterra equation \eqref{eqn:Volterra_Equation_Short}. We now use this to derive the JT equation for $\Sigma_1$. First, let us work at a fixed $k_y$ and $R_*$, so $\ell$ is fixed and we drop the superscript on $B(k_R,t)$. Then we will find it useful to write \eqref{eqn:Sigma_1_complicated} in the compact form
\begin{equation}
    {\Sigma}_1(k_R,t) = S(k_R)\me^{i\mu(k_R,t)}B(k_R, t),
    \label{eqn:Sigma_in_terms_of_B}
\end{equation}
where $S(k_R)=-(2\pi G)^{-1} u(k_R)(k_R^2+k_y^2)^{1/2}$ and $\mu(k_R,t) = \ell\Omega(t-t_\mathrm{i})+k_RR$. Using \eqref{eqn:Sigma_in_terms_of_B} we can eliminate $B$ from the Volterra equation \eqref{eqn:Volterra_Equation_Short} in favor of $ {\Sigma}_1$, with the result 
\begin{equation}
    {\Sigma}_1(k_R, t) = {\Sigma}^\mathrm{kin}_1(k_R, t) + \int_{t_\mathrm{i}}^t 
   \md t' \fint \md k_R' \frac{S(k_R)}{S(k_R')}\me^{i[\mu(k_R,t) - \mu(k_R',t')]}\mathcal{M}[k_R,k_R',t-t']
    [{\Sigma}_1(k_R', t') + {\Sigma}^\mathrm{ext}_1(k_R', t')],
\end{equation}
where $\mathcal{M}$ is given in \eqref{eqn:Volterra_Kernel_Sheet}. The presence of the delta function in \eqref{eqn:Volterra_Kernel_Sheet} means we can easily perform the integral over $k_R'$ by evaluating everything at $k_R' = k_R-2Ak_y(t-t')$.

We now follow \cite{binney2020shearing} in focusing on a particular (time-dependent) value of $k_R$, namely $k_R=k_x(t)$ where
\begin{equation}
    k_x(t) \equiv 2Ak_yt.
\end{equation}
Then the delta function in \eqref{eqn:Volterra_Kernel_Sheet} forces $k_R' = 2Akyt'$, and so 
$\Sigma_1(k_R, t) = \Sigma_1[k_x(t),t] \equiv \Sigma_1(t)$ evolves according to
\begin{equation}
    {\Sigma}_1(t) = {\Sigma}^\mathrm{kin}_1(t) + \int_{t_\mathrm{i}}^t \kappa \, \md t' K(t,t')[{\Sigma}_1(t') + {\Sigma}^\mathrm{ext}_1(t')],
    \label{eqn:JT_Equation}
\end{equation}
where $K(t,t')$ is what \cite{binney2020shearing} calls the `JT kernel'
(not to be confused with a wavenumber):
\begin{equation}
    K(t,t') = \frac{1}{\kappa}  \frac{S(2Ak_yt)}{S(2Ak_yt')}\me^{i[\mu(2Ak_yt,\,t) - \mu(2Ak_yt',\,t')]}\mathcal{N}(2Ak_yt,\,2Ak_yt',\,t-t'),
    \label{eqn:Proto_JT_Kernel}
\end{equation}
where $\mathcal{N}$ is the part of $\mathcal{M}$ that precedes the delta function.

Putting the explicit expressions for $S$, $\mu$ and $\mathcal{N}$ back in to \eqref{eqn:Proto_JT_Kernel} we get 
\begin{equation}
    K(t,t') \equiv 
    -   \frac{L}{2\pi} \frac{8\pi^3 g_0 \, a^2 R B}{\mathcal{E}\kappa}  \
    \vert u[k_x(t)]\vert^2  \sqrt{\frac{1+4A^2t^2}{1+4A'^2t'^2}}
      K_RK_R' \,
    {\sin \theta} \,\me^{-(\mathcal{K}  a)^2/4 }.
    \label{eqn:JT_Kernel_Complicated}
\end{equation}
where the factors $K_R$ and $K_R'$ (equation \eqref{eqn:modified_k}),  $\theta$ (equation \eqref{eqn:time_dependent_angle}) and $\mathcal{K}$ (equation \eqref{eqn:calK}) are evaluated at $k_R=2Ak_yt$,  $k_R' = 2Ak_yt'$ and $\tau = t-t'$. Finally, we substitute \eqref{eqn:curlyE_WKB} for $\mathcal{E}$, replace $g_0$ with
\begin{equation}
    g_0 = -\frac{\kappa \Sigma}{4\pi \sigma^2B}
    \label{eqn:Surface_Density_g0}
\end{equation}
(see equations (16)-(17) of \citealt{binney2020shearing}), and use the definition \eqref{eqn:critical_wavenumber} of the critical wavenumber $k_\mathrm{crit}$, and the fact that $a^2=2\sigma^2/\kappa^2$. The result is 
\begin{equation}
    K(t,t') =  \frac{1}{\sqrt{1+4A^2t'^2}} \frac{4\kappa}{k_\mathrm{crit}} \bm{x}\cdot\bm{y}\,\me^{-2\sigma^2 \vert \bm{y}\vert^2},
    \label{eqn:JT_Kernel}
\end{equation}
where we defined the vectors 
\begin{equation}
    \bm{x}(t) \equiv 
    \frac{1}{a\vert k_y\vert}
    \begin{pmatrix}
        K_R a/2 \\ 0
    \end{pmatrix}, \,\,\,\,\,\,\,\,\,\,\,\,\,\,\,\mathrm{and} \,\,\,\,\,\,\,\,\,\,\,\,\,\,\,   \bm{y}(t,t') \equiv 
    \frac{1}{\sqrt{2}\sigma}
    \begin{pmatrix}
        (K_R'a/2)\sin\theta  \\ -  K_Ra/2 + (K_R'a/2)\cos \theta
    \end{pmatrix}.
\end{equation}
We verified that the vectors $\bm{x}$ and $\bm{y}$ are just rotated versions of Binney's vectors $\bm{b}$ and $\bm{c}$; in particular $\vert \bm{y} \vert^2 = \vert \bm{b} \vert^2$ and $\bm{x}(t)\cdot\bm{y}(t,t') = \bm{c}(t')\cdot\bm{b}(t,t')$.\footnote{In the definition of $\bm{b}$ (equation (47) of \citealt{binney2020shearing}), the $k_y$ should strictly be $\vert k_y \vert$ (he implicitly assumed $k_y>0$).} Thus, \eqref{eqn:JT_Kernel} is exactly the JT kernel that \cite{binney2020shearing} derived --- see his equations (52) and (56) --- starting from the equations of the shearing sheet.

Finally, we note that if we had worked with the Dehnen-corrected Volterra kernel \eqref{eqn:Volterra_Kernel_Sheet_Dehnen} rather than the shearing sheet kernel \eqref{eqn:Volterra_Kernel_Sheet}, the above analysis would have been the same up to equation \eqref{eqn:Proto_JT_Kernel}. Then we would again get equation \eqref{eqn:JT_Kernel_Complicated} except for the replacement
\begin{equation}
    {\sin \theta} \,\me^{-(\mathcal{K}  a)^2/4 } \to \frac{{\sin \theta}}{\zeta^2} \me^{-(\mathcal{K}  a)^2/(4\zeta)},
\end{equation}
where $\zeta$ is given in \eqref{eqn:Dehnen_phase} and we set $\tau = t-t'$. Thus, the Dehnen-corrected version of the JT kernel \eqref{eqn:JT_Kernel} reads 
\begin{equation}
    K(t,t') =  \frac{1}{\sqrt{1+4A^2t'^2}} \frac{4\kappa}{k_\mathrm{crit}}   \frac{\bm{x}\cdot\bm{y}}{\zeta^2}\,\me^{-2\sigma^2 \vert \bm{y}\vert^2/\zeta}.
\end{equation}

\bibliography{sample}{}
\bibliographystyle{aasjournal}

\end{document}